\documentclass[preprint,nofootinbib,aps,superscriptaddress,eqsecnum]{revtex4-1}

\usepackage{graphicx}
\usepackage{mathrsfs}
\usepackage{soul}
\usepackage{natbib}
\usepackage{enumerate}
\usepackage{amsmath,amssymb}
\usepackage{slashed}
\usepackage{amssymb,amsmath,subfigure,xcolor}

\newcommand{\lapprox}{%
\mathrel{%
\setbox0=\hbox{$<$}
\raise0.6ex\copy0\kern-\wd0
\lower0.65ex\hbox{$\sim$}
}}
\newcommand{\gapprox}{%
\mathrel{%
\setbox0=\hbox{$>$}
\raise0.6ex\copy0\kern-\wd0
\lower0.65ex\hbox{$\sim$}
}}

\newcommand{\ba}{\begin{array}}
\newcommand{\ea}{\end{array}}
\newcommand{\bd}{\begin{displaymath}}
\newcommand{\ed}{\end{displaymath}}
\newcommand{\beq}{\begin{equation}}
\newcommand{\eeq}{\end{equation}}
\newcommand{\bea}{\begin{eqnarray}}
\newcommand{\eea}{\end{eqnarray}}

\newcommand{\nn}{\nonumber}

\def\ie{ {\em i.e.,\ }}

\def\a{\alpha}

\def\b{\beta}
\def\k{\kappa}
\def\g{\gamma}

\def\l{\lambda}
\def\m{\mu}
\def\n{\nu}

\def\q2 {q^2}

\def\bt{\begin{table}}
\def\et{\end{table}}

\catcode`@=11 % This allows us to modify PLAIN macros.
\def \gsim{\mathrel{\mathpalette\@versim>}}
\def \lsim{\mathrel{\mathpalette\@versim<}}
\def \@versim#1#2{\lower0.4ex\vbox{\baselineskip\z@skip\lineskip\z@skip
     \lineskiplimit\z@\ialign{$\m@th#1\hfil##\hfil$%
     \crcr#2\crcr\sim\crcr}}}
\catcode`@=12 % at signs are no longer letters

\begin{document}

\renewcommand*{\thefootnote}{\fnsymbol{footnote}}

\begin{center}

{\large\bf Alignment Limit in 2HDM: Robustness put to test}\\[15mm] 
Siddhartha Karmakar\footnote{E-mail: phd1401251010@iiti.ac.in} and Subhendu
Rakshit\footnote{E-mail: rakshit@iiti.ac.in} 
\\[2mm]

{\em Discipline of Physics, Indian Institute of Technology Indore,\\
 Khandwa Road, Simrol, Indore - 453\,552, India}
\\[20mm]
\end{center}

\begin{abstract} 
\vskip 20pt
In a two-Higgs-doublet model~(2HDM), at the vicinity of the alignment limit, the extra contributions to the couplings of the SM-like Higgs with other particles can be subdominant to the same coming from the six dimensional operators.
In this context, we revisit the alignment limit itself. It is investigated to what extent these operators can mask the actual alignment in a 2HDM. 
The bosonic operators which rescale the Higgs kinetic terms can lead to substantial change in the parameter space of the model.
We find that some other bosonic operators, which are severely constrained from the electroweak precision tests, can also modify the parameter space of 2HDM due to their anomalous momentum structures. 
A particular kind of Little Higgs model is explored as an example of 2HDM effective field theory in connection with 2HDM alignment. 
Choosing a suitable benchmark point in a Type-II 2HDM, we  highlight the possibility that the exact alignment limit is ruled out at 95\%~CL in presence of such operators.
\end{abstract}
 \vskip 1 true cm
 \pacs {}
\maketitle
\section{Introduction}
\label{intro}
After discovery of the Higgs boson~\cite{Aad:2012tfa,Chatrchyan:2012xdj}, the last missing piece of Standard Model~(SM) particle spectrum, the key challenge lies in searching physics beyond the SM. As far as the scalar sector is concerned, the two-Higgs-doublet model is the most studied extension of the SM and of immense importance considering the ongoing searches of new scalar particles in LHC. The measurements of the signal strengths of the SM-like Higgs boson are in quite good agreement with the SM predictions. As a result, the 2HDM is pushed close to the so-called `alignment limit'~\cite{Gunion:2002zf,Carena:2013ooa,Bernon:2015qea,Bernon:2015wef}. The existence of new physics beyond 2HDM is also possible, making it an interesting question to ask whether such new physics is capable of causing an apparent departure from `true' alignment in 2HDM.  

An exact alignment can be achieved by demanding the Applequist-Carrazone decoupling of the new scalars in 2HDM~\cite{Gunion:2002zf}. Alignment without decoupling~\cite{Gunion:2002zf,Carena:2013ooa,Bernon:2015qea,Bernon:2015wef,Delgado:2013zfa,Haber:2013mia,Craig:2013hca,Asner:2013psa,Bernon:2014nxa,Carena:2014nza} is a more interesting scenario because it allows for the existence of exotic scalar particles even within the reach of LHC. Though, in terms of 2HDM parameters this scenario is quite fine-tuned. Keeping this in mind, to encode the effects of new physics beyond 2HDM, we adhere to the language of 2HDM effective field theory~(2HDMEFT), which assumes both the Higgs doublets to be the low-energy fields, to investigate possible deviation from the alignment limit.  

%The attempt to write down the complete basis of 2HDMEFT has
%been done only recently~\cite{Karmakar:2017yek,Crivellin:2016ihg}.

The complete basis of operators up to dimension six in 2HDMEFT has been presented only recently~\cite{Karmakar:2017yek}. 
An earlier attempt for the same was made in ref.~\cite{Crivellin:2016ihg}.
The basis of ref.~\cite{Karmakar:2017yek} is motivated by the SILH basis~\cite{Giudice:2007fh} of SMEFT, whereas ref.~\cite{Crivellin:2016ihg} follows the Warsaw basis~\cite{Grzadkowski:2010es}. There are 126 six-dimensional operators in 2HDMEFT under the assumption of $CP$-, $B$- and $L$-conservations compared to 53 in Standard Model effective field theory~(SMEFT)~\cite{Grzadkowski:2010es}. 38 of the six-dimensional operators in 2HDMEFT are $Z_2$-violating.

One of the key observations of ref.~\cite{Karmakar:2017yek} was that the contribution of the 6-dim operators to decay width of the SM-like Higgs boson can supersede the extra contribution due to 2HDM at tree-level compared to SM. Such effects can modify the signal strengths of the SM-like Higgs boson under the framework of 2HDMEFT. It is worth exploring whether such operators are capable of masking the `true alignment' of the 2HDM. In other words, the allowed parameter space of the model can be significantly altered due to the  presence of the 6-dim operators when confronted with the measured signal strengths involving the SM-like Higgs boson. In this paper, we concentrate on the effects of the bosonic operators of 2HDMEFT and the confusion they can lead to in determining   the deviations from the alignment limit.      

In Section~\ref{operators} we introduce the 2HDM Lagrangian and set up the theoretical ground by discussing the way the 6-dim operators affect the couplings of the CP-even neutral scalars. The existing bounds on the 2HDM parameter space and the choice of parameters relevant for the present work have been discussed in Section~\ref{three}. In Section~\ref{section4} we illustrate the effects of the 6-dim operators on the alignment limit of 2HDM. We summarise and eventually conclude in Section~\ref{summary}.

\section{Relevant bosonic operators}
\label{operators}
The theoretical motivation of the 2HDM is manyfold. For example, a second Higgs doublet appears in the supersymmetric extensions of SM. Even without a supersymmetric origin, a general 2HDM has been deployed to address issues pertaining to electroweak baryogenesis~\cite{Dorsch:2013wja,Cline:2011mm}, certain flavour anomalies~\cite{Crivellin:2012ye,Crivellin:2015hha}, certain DM models etc. Moreover, many BSM models predict the existence of other particles along with a second Higgs doublet, for example, supersymmetric models~\cite{Djouadi:2015jea,Christensen:2013dra,Chen:2013jvg}, composite 2HDMs~\cite{Mrazek:2011iu}, Little Higgs models~\cite{Gopalakrishna:2015dkt,Brown:2010ke,Schmaltz:2010ac}, composite Inert doublet models~\cite{Fonseca:2015gva,Carmona:2015haa} etc. Such models can be realised as examples of 2HDMEFT~\cite{DiazCruz:2001tn,Crivellin:2016ihg,Karmakar:2017yek} where all the degrees of freedom except the two Higgs doublets are decoupled from the mass spectrum.

We define the two scalar doublets, following the notation of ref.~\cite{Karmakar:2017yek} as,
\begin{equation} 
\varphi_{I} =  \left(
\begin{array}{c}
\phi_{I}^{+}\\
\frac{1}{\sqrt{2}}(v_{I} + \rho_{I}) + i\, \eta_{I}\\
\end{array}
\right).
\end{equation}

Before spontaneous symmetry breaking~(SSB), the tree-level 2HDM Lagrangian augmented with 6-dim operators assumes the form, 
\bea
\label{lagrangian}
&&\mathcal{L} = \mathcal{L}_{kin} + \mathcal{L}_{yuk} - V(\varphi_{1},\varphi_{2}) + \mathcal{L}_{6},
\eea
where, 
\bea
\label{lagrangianterms}
\mathcal{L}_{kin} &=&  -\frac{1}{4} \sum_{X = {G^{a}},W^{i},B} X_{\m\n} X^{\m\n} + \sum_{I = 1,2} |D_{\m} \varphi_{I}|^2 + \sum_{\psi = Q,L,u,d,l} \bar{\psi} i \slashed{D} \psi,\nn\\
\mathcal{L}_{yuk} &=& \sum_{I=1,2} Y^{e}_I \, \bar{l} \,e \varphi_{I} + \sum_{I=1,2} Y^{d}_I \, \bar{q} \, d \varphi_{I} + \sum_{I=1,2} Y^{u}_I \, \bar{q} \, u \tilde{\varphi}_{I}, \nn\\
V(\varphi_{1},\varphi_{2}) &=& m_{11}^2 |\varphi_{1}|^2
 + m_{22}^2 |\varphi_{2}|^2 - ( \m^2 \varphi_{1}^{\dagger} \varphi_{2} +  h.c.) + \lambda_1 |\varphi_{1}|^4 + \lambda_2 |\varphi_{2}|^4 + \lambda_{3} |\varphi_{1}|^2 |\varphi_{2}|^2 \nn \\
  &+&  \lambda_4 |\varphi_{1}^{\dagger} \varphi_{2}|^2 + \Big(\Big( \frac{\l_5}{2} \varphi_{1}^{\dagger} \varphi_{2} + \l_6 |\varphi_{1}|^2 + \l_7 |\varphi_{2}|^2\Big)\varphi_{1}^{\dagger} \varphi_{2}  + h.c.\Big), \nn\\
  \mathcal{L}_{6} &=& \sum_{i} c_i O_{i}/\Lambda^2. 
\eea
In eqn.~(\ref{lagrangianterms}), $O_{i}$ are the 6-dim operators whose effects we are interested in, $c_i$ being their corresponding Wilson coefficients. $\Lambda$ is the scale of new physics beyond 2HDM. $\l_{6,7}$ are the so-called hard $Z_2$-violating terms in the 2HDM potential.

In 2HDM, after SSB, extra physical scalar fields appear in the mass spectrum. The charge-neutral physical scalars, $h$ and $H$ are two orthogonal combinations of the unphysical fields $\rho_{1,2}$. Similarly, the charged scalars $H^{\pm}$, and neutral pseudoscalar $A$, are certain combinations of $\phi_{1,2}^{\pm}$ and $\eta_{1,2}$ respectively. The Goldstone modes which get absorbed as the longitudinal degrees of freedom of the $W^{\pm}$ and $Z$ bosons are orthogonal to the mass eigenstates  $H^{\pm}$ and $A$ respectively. The rotation required to transform the unphysical fields $\phi_{1,2}^{\pm}$ and $\eta_{1,2}$ into mass eigenstates is denoted by the angle $\beta$ = $\tan^{-1} (v_2/v_1)$. Similarly, the rotation required to mass-diagonalise the neutral scalars is parametrised by the angle $\alpha$~\cite{Carena:2013ooa}: 
\bea
\alpha &=& \sin^{-1} \Big[\frac{\mathcal{M}_{12\rho}^2}{\sqrt{(\mathcal{M}_{12\rho}^2)^2 + (\mathcal{M}_{11\rho}^2 - m_{h}^2)^2}}\Big],
\eea
where $\mathcal{M}^2_{ij\rho}$ is the $ij$-th element of the mass-squared matrix for the fields $\rho_1$ and $\rho_2$. It is due to these two rotations in the scalar sector, the couplings of the SM-like Higgs boson, $h$ to the massive vector bosons are scaled compared to the SM scenario. In fact all the couplings of SM-like Higgs boson in 2HDM are their SM counterparts times a coupling multiplier. The coupling multipliers for the interaction of the $CP$-even neutral scalars,\ie $h$ and $H$, to a pair of massive gauge bosons are given by: 
\bea
\label{couplmulthvv}
\k_{hVV} = \sin (\b - \a), \hspace{15pt} \k_{HVV} = \cos (\b - \a),
\eea
with $\k_{s XX} = g_{s XX}/g_{hXX}^{SM}$, where $g_{hXX}^{SM}$ is the coupling of Higgs boson to the species $X$ in SM and $g_{sXX}$ is the coupling of $CP$-even neutral scalars $s = h, H$, to the species $X$ in 2HDM. In the present paper we work under the assumption of $m_{H} > m_h$. However, the alternative scenario with $m_H < m_{h}$ has also been investigated in literature~\cite{Bernon:2015wef}.

In order to suppress the Higgs-mediated flavour-changing neutral current~(FCNC), it is important to make sure that no fermion gets mass from both the doublets~\cite{Glashow:1976nt}. This is ensured by the assumption of a $Z_2$-symmetry under which: $\varphi_1 \rightarrow \varphi_1$ and $\varphi_2 \rightarrow -\varphi_2$. There can be four ways in which the assignment of the $Z_2$-charges to the fermions can be performed, leading to four different kinds of Yukawa interactions, Type-I, II, Lepton-specific and Flipped~\cite{Branco:2011iw,Aoki:2009ha}. In this paper, we concentrate on only the first two kinds, which are the most studied ones in literature. For Type-I 2HDM, all the fermions, the $u$-type and $d$-type quarks and the charged leptons, get mass from the second doublet. For Type-II, $u$-type quarks get their masses from $\varphi_2$, while $\varphi_1$ provides masses to $d$-type quarks and the charged leptons. Similar to $hVV$ couplings, the couplings of the SM-like Higgs to fermion pairs also get modified in 2HDM. The coupling multipliers for the interactions of the neutral scalars with a pair of SM fermions are given as: 
\bea
\label{coupmulthff}
\text{Type-I}:&& \hspace{30pt}
\begin{cases}
\kappa_{huu} = \kappa_{hdd} = \kappa_{hll} = \sin (\beta - \alpha) +  \cos (\beta - \alpha)/\tan \beta, \\
\kappa_{Huu} = \kappa_{Hdd} = \kappa_{Hll} = \cos (\beta - \alpha) -  \sin (\beta - \alpha)/\tan \beta, \\
\kappa_{Auu} = \cot \b, \kappa_{Add} = \kappa_{All} = -\cot \b,\\
\end{cases}\\
\text{Type-II}:&& \hspace{30pt}
\begin{cases}
\kappa_{huu} =  \sin (\beta - \alpha) +  \cos (\beta - \alpha)/\tan \beta,\\
\kappa_{hdd} = \kappa_{hll} = \sin (\beta - \alpha) -  \cos (\beta - \alpha) \tan \beta,\\
\kappa_{Huu} =  \cos (\beta - \alpha) -  \sin (\beta - \alpha)/\tan \beta,\\
\kappa_{Hdd} = \kappa_{Hll} = \cos (\beta - \alpha) + \sin (\beta - \alpha) \tan \beta,\\
\kappa_{Auu} = \cot \b, \kappa_{Add} = \kappa_{All} = \tan \b.
\end{cases}\nn
\eea 

Due to such modifications in the couplings, the production and decay rates of the SM-like Higgs boson in 2HDM differ from those in SM. After the discovery of the 125 GeV Higgs boson, whose properties are in quite good agreement with that of the SM Higgs boson, the experimentally allowed parameter space of 2HDM gets constrained. It can be seen from eqns.~(\ref{couplmulthvv}) and (\ref{coupmulthff}) that all the coupling multipliers are functions of $\cos (\b-\a)$ and $\tan \b$. Moreover, in the limit $\cos (\b-\a) \rightarrow 0$ all the couplings reduce to corresponding SM values. This is known as the `alignment limit' of 2HDM and will be discussed in detail in Section~\ref{subs:sigstr}. If certain additional symmetries are imposed, alignment limit can be achieved as a natural consequence in models with extra Higgs doublets~\cite{Dev:2014yca,Pramanick:2017wry}. Also, the case of a 2HDM emerging as a low-energy effective theory of a supersymmetric scenario has been explored~\cite{Benakli:2018vqz}. 

These coupling multipliers get further changed in presence of the higher-dimensional operators. So the production rates as well as the decay width of the SM-like Higgs boson get modified. As mentioned earlier, such changes can be  larger compared to the extra contribution in 2HDM at tree-level~\cite{Karmakar:2017yek}. This in turn affects the extraction of bounds on the parameter space in 2HDM which can significantly modify the alignment limit itself. 

In this paper, we intend to study the effects of the higher-dimensional operators on the alignment limit in detail. Here we concentrate on the effects of the bosonic 6-dim operators. The impact of the fermionic operator will be discussed elsewhere. The complete set of  bosonic operators in case of 2HDM can be found in ref.~\cite{Karmakar:2017yek}. The power counting based on naive dimensional analysis~(NDA) renders some of the operators more  suppressed than the rest, thus making some operators more significant when it comes to phenomenological analyses.   

The operators under consideration are of types $\varphi^4 D^2$, $\varphi^2 D^2 X$, $\varphi^2 X^2$ and $\varphi^6$. These are the only bosonic operators which involve the scalar fields, hence important for Higgs physics. As we go on we will see that the first two types of operators are relevant for our discussion.

\noindent $\bullet$ ${\bf \varphi^4 D^2}$

These operators lead to the rescaling of the kinetic terms of all the Higgs fields, sans the charged scalars. Such effects should be taken care of by appropriate field redefinitions, which lead to the scaling of the couplings of the SM-like Higgs. Some of these operators contribute to the $T$-parameter. In order to suppress such contributions, it is a common practice, for most of the UV-complete models that lead to 2HDMEFT at low energies, to assume the existence of an unbroken global $SO(4)$ symmetry, under which the two Higgs doublets transform as bidoublets~\cite{Mrazek:2011iu}. As these operators contribute to the $T$-parameter at tree-level, the corresponding Wilson coefficients are constrained at $\sim \mathcal{O}(10^{-3})$~\cite{Karmakar:2017yek,Pomarol:2013zra,Falkowski:2013dza}. Thus, these operators lead to insignificant changes in the decay widths of the SM-like Higgs in various channels and we neglect them. So, we are left with the following operators~\cite{Karmakar:2017yek},
\bea
\label{ops}
O_{H1} &=& (\partial_{\m}|\varphi_1|^2)^2, \hspace{10pt} O_{H2} = (\partial_{\m}|\varphi_2|^2)^2,\hspace{10pt} O_{H12} = (\partial_{\m}(\varphi_1^{\dagger} \varphi_2 + h.c.))^2, \\
O_{H1H2} &=& \partial_{\m}|\varphi_1|^2 \partial^{\m}|\varphi_2|^2, O_{H1H12} = \partial_{\m}|\varphi_1|^2\partial^{\m}(\varphi_1^{\dagger} \varphi_2 + h.c.), O_{H2H12} = \partial_{\m}|\varphi_2|^2\partial^{\m}(\varphi_1^{\dagger} \varphi_2 + h.c.). \nn
\eea
Operators $O_{H1H12}$ and $O_{H2H12}$ are odd under the $Z_2$-symmetry, whereas the rest are even.
In presence of these higher dimensional operators, the angle $\b$ still diagonalises the charged and $CP$-odd scalars as in 2HDM at the tree-level~\cite{Karmakar:2017yek}.
However, the $CP$-even neutral scalars $\rho_{1,2}$ can no longer be diagonalised by the mixing angle $\alpha$ as these scalar fields have to be redefined in the following way in order to achieve canonically normalised kinetic terms~\cite{Karmakar:2017yek}:
\bea
\label{normalisation}
\rho_1 &\rightarrow & \rho_1 \Big(1 - \frac{\Delta_{11\rho}}{4f^2}\Big) - \rho_2 \frac{\Delta_{12\rho}}{8f^2}, \nn\\
\rho_2 &\rightarrow & \rho_2 \Big(1 - \frac{\Delta_{22\rho}}{4f^2}\Big) - \rho_1\frac{\Delta_{12\rho}}{8f^2},
\eea    
where $\Delta_{ij\rho}$ are defined in Appendix~\ref{appendix1}.
This leads to the redefinition of the physical CP-even scalar fields as:
\bea
\left(
\begin{array}{c}
H\\
h\\
\end{array}\right)
\rightarrow \begin{bmatrix}
    c_{\a}        & s_{\a}  \\
    -s_{\a}      & c_{\a}    \\
\end{bmatrix}\times
\begin{bmatrix}
    1 - \frac{\Delta_{11\rho}}{4f^2}        & -\frac{\Delta_{12\rho}}{8f^2}  \\
    -\frac{\Delta_{12\rho}}{8f^2}       & 1 - \frac{\Delta_{22\rho}}{4f^2}    \\
\end{bmatrix} \times
\begin{bmatrix}
    c_{\a}        & -s_{\a}  \\
    s_{\a}      & c_{\a}    \\
\end{bmatrix}
\left(
\begin{array}{c}
H\\
h\\
\end{array}\right)
\eea
or, 
\bea
\label{hfieldred}
h & \rightarrow & (1-x_1)h+  y H,\nn\\
H & \rightarrow & (1-x_2)H+  y h
\eea
$x_1$, $x_2$ and $y$ are functions of the Wilson coefficients of the higher dimensional operators as given in Appendix~\ref{appendix1}. In our notation, $c_{\theta} \equiv \cos \theta$, $s_{\theta} \equiv \sin \theta$ and $t_{\theta} \equiv \tan \theta$. Due to this redefinition of fields, couplings of both the CP-even neutral scalars to vector bosons and fermions get modified compared to 2HDM at the tree-level. The coupling multipliers of the SM-like Higgs boson and of the other neutral scalar $H$ are modified as follows:
\bea
\label{scalemult}
\kappa_{hff}^{\prime} &=& (1- x_1)\kappa_{hff}+ y \kappa_{Hff},\nn\\
\kappa_{hVV}^{\prime} &=& (1 - x_1) \sin (\b-\a) + y \cos (\b-\a), \nn\\ 
\kappa_{Hff}^{\prime} &=& (1- x_2)\kappa_{Hff}+ y \kappa_{hff},\nn\\
\kappa_{HVV}^{\prime} &=& (1 - x_2) \cos (\b-\a) + y \sin (\b-\a),
\eea
 where $V = W,Z$. These modified coupling multipliers reduce to that in 2HDM at tree-level with $f \rightarrow \infty$, as expected. The reason that the coupling multipliers of the $hWW$ and $hZZ$ vertices are the same lies in the fact that we have ignored the $T$-parameter violating operators in this paper. In presence of these operators the coupling multiplier for $hZZ$ vertex gets additional contributions compared to the $hWW$ vertex. As we neglect these operators, the $AVV$ and $Aff$ vertices do not modify compared to 2HDM at tree-level.

It is interesting to note that these operators lead to additional momentum-dependent terms in the triple scalar vertices. However, in this paper we are mainly interested in the bounds coming from the signal strengths of the SM-like Higgs boson. The detailed phenomenology of the exotic scalars under the framework of 2HDMEFT will be addressed elsewhere.  

%\vspace{10pt}
%\pagebreak

\noindent $\bullet$ ${\bf \varphi^2 D^2 X}$ {\bf and} ${\bf \varphi^2 X^2}$

There are 12 different operators of type $\varphi^2 D^2 X$, namely $O_{Bij}$, $O_{Wij}$, $O_{\varphi Bij}$ and $O_{\varphi Wij}$ in our basis defined as~\cite{Karmakar:2017yek}:  
\begin{eqnarray}
\label{definitns}
O_{Bij} &=& \frac{ig^{\prime}}{2} (\varphi^{\dagger}_i \overset\leftrightarrow{D_{\m}} \varphi_j) D_{\nu} B^{\m\n}, \hspace{10pt}
O_{Wij} = \frac{ig}{2} (\varphi^{\dagger}_i \vec{\sigma} \overset\leftrightarrow{D_{\m}} \varphi_j) D_{\nu} \vec{W}^{\m\n}, \nn\\
O_{\varphi Bij} &=& ig^{\prime}  (D_{\mu} \varphi_{i}^{\dagger} D_{\nu} \varphi_{j}) B^{\m\n}, \hspace{10pt}
O_{\varphi Wij} = ig  (D_{\mu} \varphi_{i}^{\dagger} \vec{\sigma} D_{\nu} \varphi_{j}) \vec{W}^{\m\n},
\end{eqnarray}
with, $i,j = 1,2$. These types of operators contribute to observables, pertaining to precision tests and SM-like Higgs phenomenology. For example, $O_{Bij}, O_{Wij}$ contribute to $S$-parameter and $O_{\varphi Bij}, O_{\varphi Wij}$, $O_{Wij}$ to the anomalous triple gauge boson vertices. The precision observables have been precisely measured at LEP and thus constrain the Wilson coefficients of these operators around $\mathcal{O}(10^{-2})$~\cite{LEPEWWG}. Inclusion of Higgs data to the fits makes such constraints even more stringent~\cite{Karmakar:2017yek,Pomarol:2013zra,Falkowski:2013dza}.  
 
There are 3 operators of type $\varphi^2 X^2$ for $X=B$ given as:
\bea
O_{BBij} &=&  g^{\prime 2} (\varphi^{\dagger}_{i} \varphi_{j})B_{\m\n}B^{\m\n}.
\eea
The most stringent constraint on the Wilson coefficients of these operators come from the measurement of the decay width $h \rightarrow \gamma \gamma, Z \gamma$  for $O_{BBij}$. The bounds on the Wilson coefficients of $O_{BBij}$ are around $\mathcal{O}(10^{-3})$~\cite{Karmakar:2017yek,Pomarol:2013zra,Falkowski:2013dza}, making these operators insignificant for our purpose. Such an operator with $W$ bosons is not present in our basis~\cite{Karmakar:2017yek}. We have refrained from discussing the effects of $O_{GGij}$ which can be constrained from $\sigma (gg \rightarrow h)$.  
 
\pagebreak

\section{Existing constraints}
\label{three}
\noindent\subsection{Choice of 2HDM parameters}
\label{bounds}
In this paper, we work in the physical basis of the 2HDM in which the complete set of parameters $\{m_{h},\, m_{H},\, m_{A} ,\, m_{H^{\pm}},\, \cos (\beta-\alpha),\, \tan \beta,\, \m^{2},\, v\}$ describes the potential of the model. $\m^2$ is the coefficient of the $Z_2$-odd quadratic term in the 2HDM potential appearing in eqn.~(\ref{lagrangianterms}) and $v \sim 246$ GeV is the electroweak vev. In defining $\cos (\beta-\alpha)$ we use the convention as in ref.~\cite{Haber:2015pua}.
In 2HDM at tree-level, the coupling multipliers are sole functions of $\cos (\b-\a)$ and $\tan \b$. It can also be seen from eqns.~(\ref{scalemult}) and (\ref{phi2d2xmult}) that the changes in the couplings due to the 6-dim operators also depend on these two 2HDM parameters. Thus, as it was mentioned earlier, the alignment limit in 2HDM and its modifications after the inclusion of the 6-dim operators are best demonstrated on the $\cos (\beta-\alpha) - \tan \beta$ plane.
 
The hard $Z_2$-violating couplings, $\lambda_6$ and $\lambda_7$ have been set to zero for now, though it is worth mentioning that even non-zero values of $\l_6$ and $\l_7$ can be rotated away into $\l_6 = \l_7 = 0$ exploiting the reparametrisation invariance of the 2HDM Lagrangian as long as certain conditions in terms of other 2HDM parameters are fulfilled~\cite{Ginzburg:2004vp}.
On the $\cos (\b-\a) - \tan \beta$ plane, in case of non-zero values of $Br (H \rightarrow h h)$, region of lower $\tan{\beta}$ gets excluded~\cite{Dorsch:2016tab}. However, for most of our benchmark points we have chosen $m_{H} < 2 m_h$, so that the decay channel $H \rightarrow h h$ is kinematically forbidden. 

The direct bound on the mass of the charged scalars come from the measurements of LEP, which dictates $m_{H^{\pm}} \gtrsim 72$ GeV (80 GeV) for Type-I~(II) 2HDM~\cite{Abbiendi:2013hk}. LEP searches also put constraint on the sum of the masses of neutral exotic scalars, $m_{H} + m_{A} \gtrsim 209$ GeV~\cite{Schael:2006cr}. We work under the approximation of $m_{A} \sim m_{H^{\pm}}$, which ensures that the contribution to $T$-parameter at one-loop is rather small~\cite{Gunion:1989we}. As mentioned earlier, we have neglected the operators of type $\varphi^4 D^2$ which violate the custodial symmetry at tree-level as their Wilson coefficients are constrained at $\mathcal{O}(10^{-3})$. So the benchmark points we use are safe from the $T$-parameter constraint. The charged scalar mediates the process $b \rightarrow s \gamma$, which is well-measured from $Br (B \rightarrow X_{s} \gamma)$.  In case of Type-II 2HDM, this in turn leads to a bound on the charged scalar mass $m_{H^{\pm}} \gtrsim 480$ GeV, which is almost independent of $\tan \beta$~\cite{Eberhardt:2013uba,Deschamps:2009rh}. Thus, we choose the value $ m_{H^{\pm}} \sim 485$ GeV in case of Type-II 2HDM.  Such bound in case of Type-I 2HDM is not as stringent as the Type-II case. For Type-I, the constraints from the measurements of meson decays lead to $m_{H^{\pm}} \gtrsim 160$ GeV~\cite{Branco:2011iw,Mahmoudi:2009zx}.

The key decay channels of $A$ near $m_{A} \sim 400$ GeV are $A \rightarrow Z h, \tau\bar{\tau}$, which rule out regions at lower $\tan \beta$ for both Type-I and Type-II 2HDM~\cite{Dorsch:2016tab}. On the $\cos (\b-\a) - \tan \b$ plane regions with higher values of $|\cos(\b-\a)|$ get excluded from the non-observation of $A \rightarrow Z h$. However for Type-II 2HDM, $A \rightarrow \tau\bar{\tau}$ rules out region with higher values of $\tan \b$ as well.
To find out the excluded region on the parameter space we have used the bounds on heavy scalars from the measurement in different search channels which have been specifically mentioned in table~I.

The theoretical bounds of stability, perturbativity and unitarity of the S-matrix~\cite{Bhattacharyya:2015nca,Horejsi:2005da,Maalampi:1991fb} on the renormalisable 2HDM parameter space have been calculated using {\tt 2HDMC-1.7.0}~\cite{Eriksson:2009ws}. In order to ensure that the stability criteria is respected, we choose for all our benchmark scenarios,

\bea
\mu^2 = \mu_{*}^2 = 
\frac{m_h^2 (s_{\b-\a} -  c_{\b - \a} t_{\b})^2 + m_{H}^2  (c_{\b-\a} +  s_{\b - \a} t_{\b})^2}{t_{\b} (1 + t_{\b}^2)} - 50,
\eea
so that $\l_1 >0$. Perturbativity is kept under control for most of the region on $\cos (\beta - \alpha) - \tan \beta$ plane by keeping the masses of the scalars close to the electroweak symmetry breaking scale.

The $\varphi^6$ type of operators modify the constraints coming from the stability of the potential in 2HDMEFT. The 6-dim operators of type $\varphi^4 D^2$ can be constrained from the unitarity of the S-matrix. Irrespective of the values of $c_{\b-\a}$ and $\tan \beta$, operators with Wilson coefficient $\sim 1$ and new physics scale $f \sim 750$ GeV violate the unitarity condition at $\sqrt s \sim 1.5 - 2$ TeV~\cite{Kikuta:2012tf,Kikuta:2011ew}. We have chosen the values of Wilson coefficients and the scale of new physics respecting such bounds.

We have taken into account all the theoretical as well as experimental constraints on parameters for all the benchmark scenarios while working on the $\cos(\b-\a)-\tan \b$ plane. Our choice of parameters makes it easier to illustrate the effects of the 6-dim operators on the 2HDM parameter space.  

\noindent \subsection{Signal strengths}
\label{subs:sigstr}
The signal strength of Higgs boson in a particular channel is defined as,
\bea
\mu_{i}^{f} = \frac{[\sigma( i\rightarrow h)\times Br(h\rightarrow f)]}{[\sigma( i\rightarrow h)\times Br(h\rightarrow f)]_{SM}}.
\eea 
The measurements of Higgs signal strengths imply that the experimental values of Higgs decay widths are in quite good agreement with the values predicted by the SM. The key constraints on the $\cos (\beta - \alpha) - \tan \beta$ plane of 2HDM parameter space come from the experimentally allowed range of the  following signal strengths at $95\%$ CL at $\sqrt{s} = 8, 13$ TeV~\cite{Khachatryan:2016vau,ATLAS:2017myr,ATLAS:2017ovn,Aaboud:2017vzb,ATLAS:2016gld,Aaboud:2017xsd,CMS:2017rli,Sirunyan:2017exp,CMS:2017pzi,Sirunyan:2017khh,CMS:2016mmc}. Among these, the most relevant ones in the context of this paper are listed in table~I. Also, we mention the LHC exotic scalar searches in various channels which we use in order to put bounds on the parameter space along with the references to the corresponding analysis.   

\begin{table}[h!]
\begin{center}
\begin{tabular}{c c c}
\hline
ATLAS, 13 TeV & \,\,\,\,\,\,\,\,\,\,\,\,\,\,\,\,\,\,\,\,\,\hspace{20pt} & CMS, 13 TeV\\
\hline
$\mu_{gg}^{\g\g} \in [0.42,1.18]$ & &$\mu_{gg}^{\g\g}  \in  [0.73,1.49]$ \\
$\mu_{gg}^{ZZ} \in  [0.63,1.59]$ & &$\mu_{gg}^{ZZ} \in [0.76,1.64]$\\
$\mu_{Vh}^{bb} \in  [0.42,1.98]$ & & $\mu_{gg}^{WW} \in  [0.48,1.56]$\\
\hline
$gg/b\bar{b} \rightarrow A/H \rightarrow \tau\bar{\tau}$& & \cite{Aaboud:2017sjh}\\
$gg/b\bar{b}  \rightarrow A/H \rightarrow \gamma\gamma$& & \cite{Aaboud:2017yyg} \\
$gg/b\bar{b} \rightarrow H \rightarrow hh$& & \cite{ATLAS:2016ixk} \\
$gg \rightarrow H \rightarrow WW$& & \cite{Aaboud:2017gsl} \\
$gg \rightarrow H \rightarrow ZZ$& & \cite{Aaboud:2017rel} \\
$gg/b\bar{b}  \rightarrow A \rightarrow Zh/ZH$& & \cite{TheATLAScollaboration:2016loc} \\
\hline
\end{tabular}
\end{center}
\label{table:data}
\caption{Signal strengths and heavy Higgs searches used in this paper.}
%\end{table}
\end{table}

We use the most stringent bounds among the values provided by ATLAS and CMS for a particular channel for Higgs signal strengths. Moreover, we choose the values corresponding to Run-I of LHC if it is more stringent than Run-II results. For example, we consider the upper bound for $\mu_{gg}^{VV}$ as 1.42~\cite{Khachatryan:2016vau} because it is more stringent than the value corresponding to Run-II.  In 2HDM, the signal strengths roughly go as: $\mu_{gg}^{VV} \sim \kappa_{htt}^2 \kappa_{hVV}^2$, for $V = W, Z$ and $\mu_{gg}^{\g\g} \sim \kappa_{htt}^4$. The signal strengths tend to unity only when all these coupling multipliers approach the unit value, which happens, as eqns.~(\ref{couplmulthvv}) and (\ref{coupmulthff}) suggest, at the alignment limit,\ie $c_{\b - \a} \rightarrow 0$. Eqn.~(\ref{coupmulthff}) further implies, for higher values of $\tan \beta$, $hbb$ coupling deviates from its SM counterpart significantly, ruling out the high-$\tan \beta$ region from the observation of $\sigma(gg \rightarrow h)$ as well as $\mu_{Vh}^{b\bar{b}}$~\cite{Craig:2013hca}. Though the situation is comparably relaxed in Type-I 2HDM where the $hbb$ coupling is less sensitive to $\tan \beta$ for $\tan \beta \gtrsim 4$, allowing quite higher values of $\cos (\beta - \alpha)$ compared to Type-II 2HDM.

Studies of 2HDM parameter space in light of the discovery of the Higgs boson at 125 GeV and future prospect of the searches of exotic scalars have been done in the literature~\cite{Bernon:2015qea,Haber:2015pua,Coleppa:2013dya,Bernon:2015wef,Craig:2015jba,Chang:2013ona,Chen:2013rba,Ferreira:2017bnx,Han:2017pfo}. The global fits on the 2HDM parameter space have been performed in ref.~\cite{Chowdhury:2015yja,Chowdhury:2017aav} taking into consideration the Higgs data, exclusion limits on the new scalars, as well as the EWPT and flavour constraints. However, performing such global fits in 2HDMEFT is beyond the scope of our present work.

\section{Alignment limit with 6-dim operators}
\label{section4}
The 6-dim operators affect the production and decay channels of the SM-like Higgs boson in non-trivial ways. The $\varphi^4 D^2$ operators redefine the Higgs fields, leading to the rescaling of the $hVV$ couplings. In contrary, $\varphi^2 D^2 X$ type of operators can modify the momentum structure of the same. Due to this, even though these operators are highly constrained from the electroweak precession observables, they can lead to significant changes in the Higgs decay widths. We determine allowed regions on the $\cos (\b-\a) - \tan \beta$ plane in presence of these 6-dim operators. To compute the decay width of the SM and exotic scalars into various channels, we have used {\tt 2HDMC-1.7.0}~\cite{Eriksson:2009ws} incorporating the modified couplings. The production cross sections for various Higgses for both the gluon fusion and $b\bar{b}$-associated production modes have been computed up to NNLO in QCD using {\tt SusHi-1.6.0}~\cite{Harlander:2012pb,Harlander:2013qxa}. 
For production cross-section of the SM-like Higgs via vector boson fusion~(VBF) and associated production with vector boson~(VH) mode we use, $\sigma / \sigma_{SM} \sim \kappa_{hVV}^2~(\kappa_{hVV}^{\prime 2})$ for 2HDM~(2HDMEFT).
We neglect the loop-level effects in the $h \rightarrow VV$ decay channels which are subleading to the tree-level contributions~\cite{Altenkamp:2017ldc}.

In the remaining part of this section we discuss the effects of the 6-dim bosonic operators under consideration on the alignment limit for the Type-I and Type-II 2HDM. We also study the case of a concrete UV-complete model, namely the little Higgs model based on the coset $SU(6)/Sp(6)$. 

\noindent \subsection{{\bf Type-I 2HDM}}
\label{subsection:3A}
\subsubsection{$\mathbf{\varphi^4}\mathbf{D^2}$}
In presence of such operators, the modified $hVV$ coupling multipliers are given in eqn.~(\ref{scalemult}). Due to such modifications the signal strength $\mu_{gg}^{VV}$ changes. The process $h \rightarrow \g\g$ is mediated by the $W^{\pm}$ boson, charged fermions and scalars, thus $\mu_{gg}^{\g\g}$ also gets modified due to the presence of these operators. 

\begin{figure}[h!]
 \begin{center}
\subfigure[]{
 \includegraphics[width=2.8in,height=2.8in, angle=0]{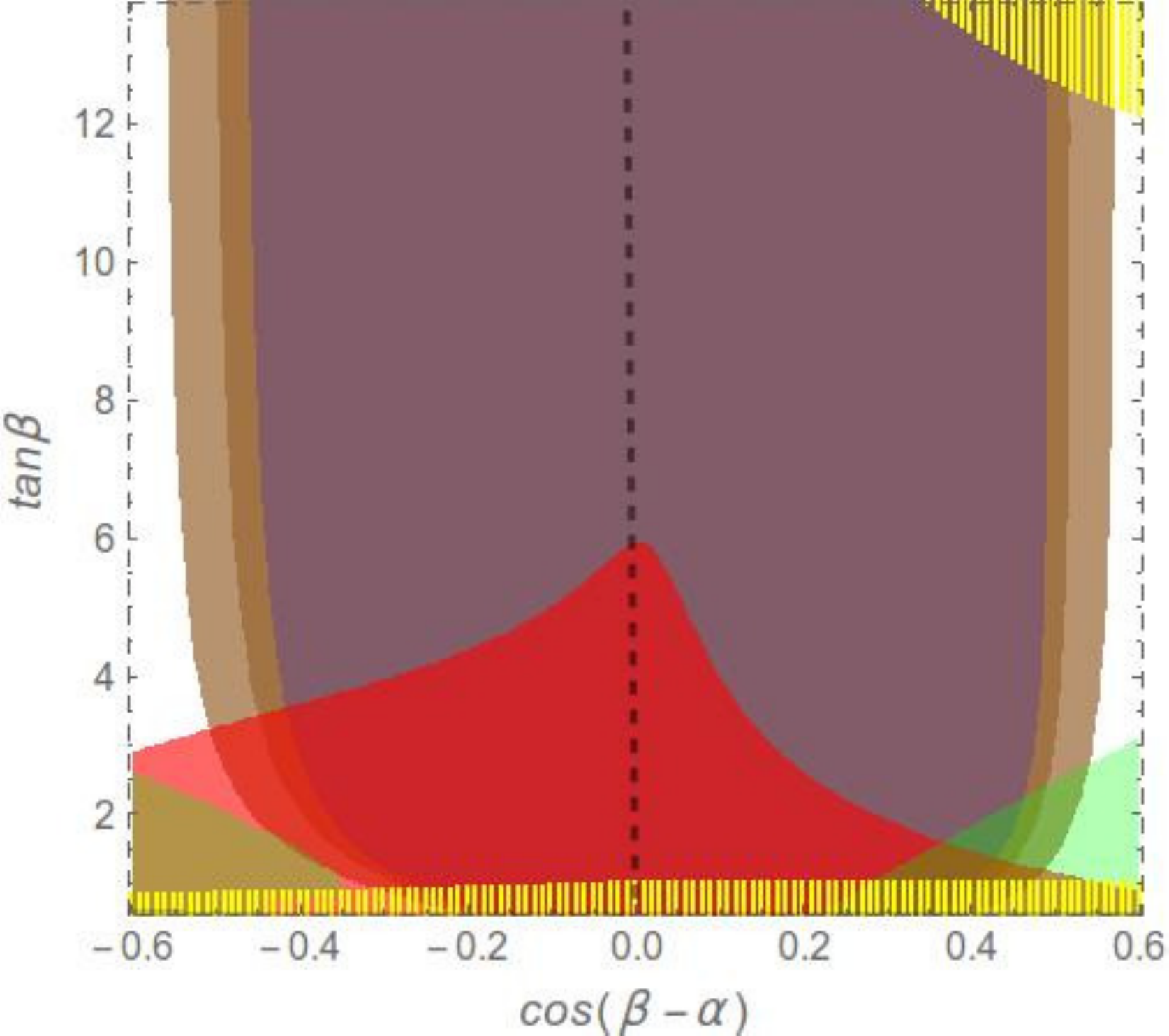}} 
 \hskip 15pt
 \subfigure[]{
 \includegraphics[width=2.8in,height=2.8in, angle=0]{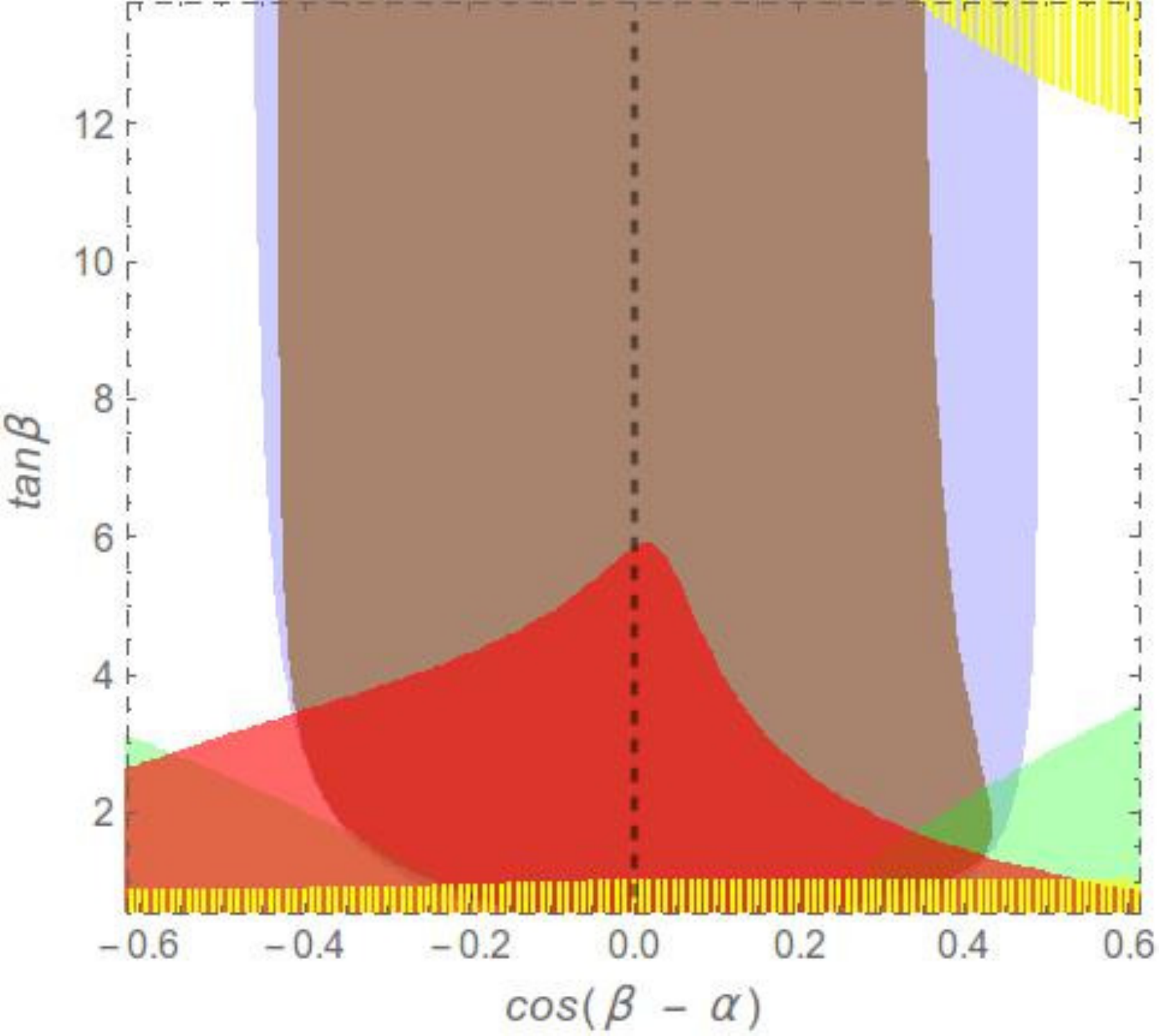}} 
 \caption{The effect of $\varphi^4 D^2$ type of operators for Type-I 2HDM: (a) BP1, (b) BP2, as described in the text. Blue and brown regions indicate areas allowed from the measurement of the signal strengths of SM-like Higgs boson, in 2HDM at tree-level and  in presence of 6-dim operators respectively. For (a), lighter and darker brown region represent BP1 with $f=1$~TeV and 2~TeV respectively. The yellow shadowed region is ruled out by theoretical criteria such as perturbativity, stability, unitarity of the S-matrix. The green and red regions are ruled out from non-observation of $A \rightarrow Z h(b\bar{b})$ and $A \rightarrow ZH(b\bar{b})$ respectively  including the effects of 6-dim operators for $f=1$~TeV for both BP1 and BP2.}
 \label{fig:phi4d2}
\end{center}
 \end{figure}
For Type-I 2HDM, the allowed region on the $\cos (\b-\a) -\tan \beta$ plane, considering the bounds on the signal strengths involving SM-like Higgs boson at $95\%$ CL, has been shown in fig.~\ref{fig:phi4d2} for the following set of 2HDM parameters:

$m_{h} = 125.7$ GeV, $m_H = 150$ GeV, 
$m_A = m_{H^{\pm}} = 400$ GeV,
$\mu^2 = \mu^2_{*}$.

To illustrate the effect of the 6-dim operators on the allowed region on this plane we choose the following benchmark points in terms of the Wilson coefficients and scale $f$,
\begin{itemize}
\item {\bf BP1 } \hspace{5pt}  $c_{H1} = c_{H2} = c_{H12} = -1$, $
 c_{H1H2} =  c_{H1H12} = c_{H2H12} = 0$, $f = 1, 2$ TeV, 
\item  {\bf BP2 } \hspace{5pt} $c_{H1} = c_{H2} = 1$, $c_{H1H12} = - c_{H2H12} = \frac{3}{2}$, $c_{H1H2} = c_{H12} = 0$, $f = 1$ TeV. 
\end{itemize}
For BP1 with $f = 1$~TeV, at $t_\b \sim 10$, the maximum allowed value of $c_{\b-\a}$ in the positive $c_{\b-\a}$ direction changes to $\sim 0.56$ from $\sim 0.48$ which is the corresponding value for 2HDM at tree-level, increasing the allowed region by $ +14\%$. In the negative region, for the same $t_{\b}$ value, maximum allowed region go from $\sim -0.45$ to $\sim -0.55$,\ie increases by $+22\%$. For BP1 with $f = 2$~TeV, at $t_\b \sim 10$, maximum allowed region in terms of $c_{\b-\a}$ go up to $\sim 0.516$ in $+$ve direction and $\sim -0.495$ in $-$ve direction, leading to respectively $+5.7\%$ and $+10\%$ change in the positive and negative $c_{\b-\a}$ directions compared to 2HDM at the tree-level respectively.  
For BP2, the maximum allowed values in the positive and negative $c_{\b-\a}$ directions become $\sim 0.36$ and $\sim -0.43$,\ie changes by $-25\%$ and $-4.4\%$ respectively.

\begin{figure}[h!]
 \begin{center}
\subfigure[]{
\includegraphics[width=2.8in,height=2.8in, angle=0]{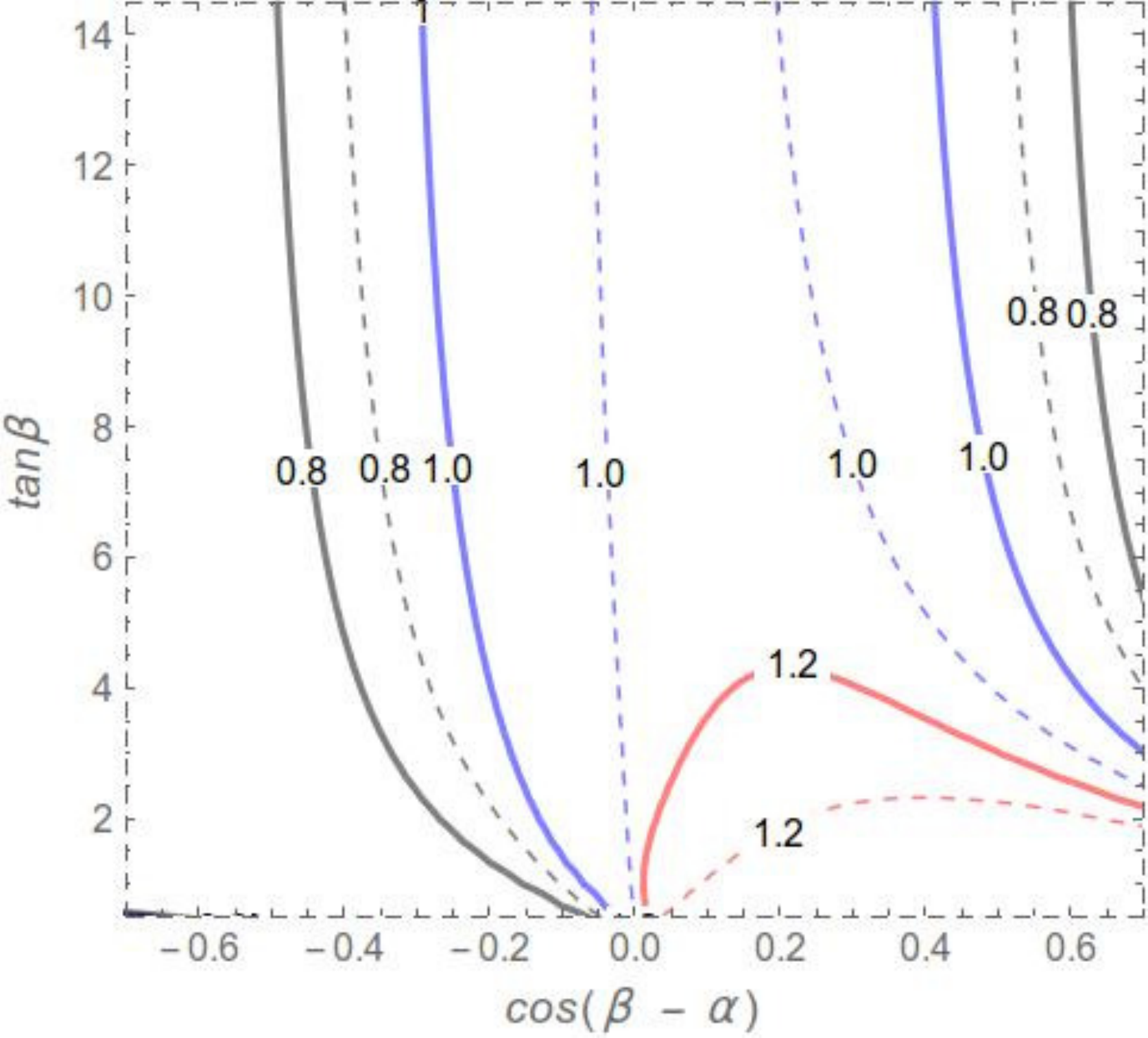}} 
 \hskip 15pt
 \subfigure[]{
 \includegraphics[width=2.8in,height=2.8in, angle=0]{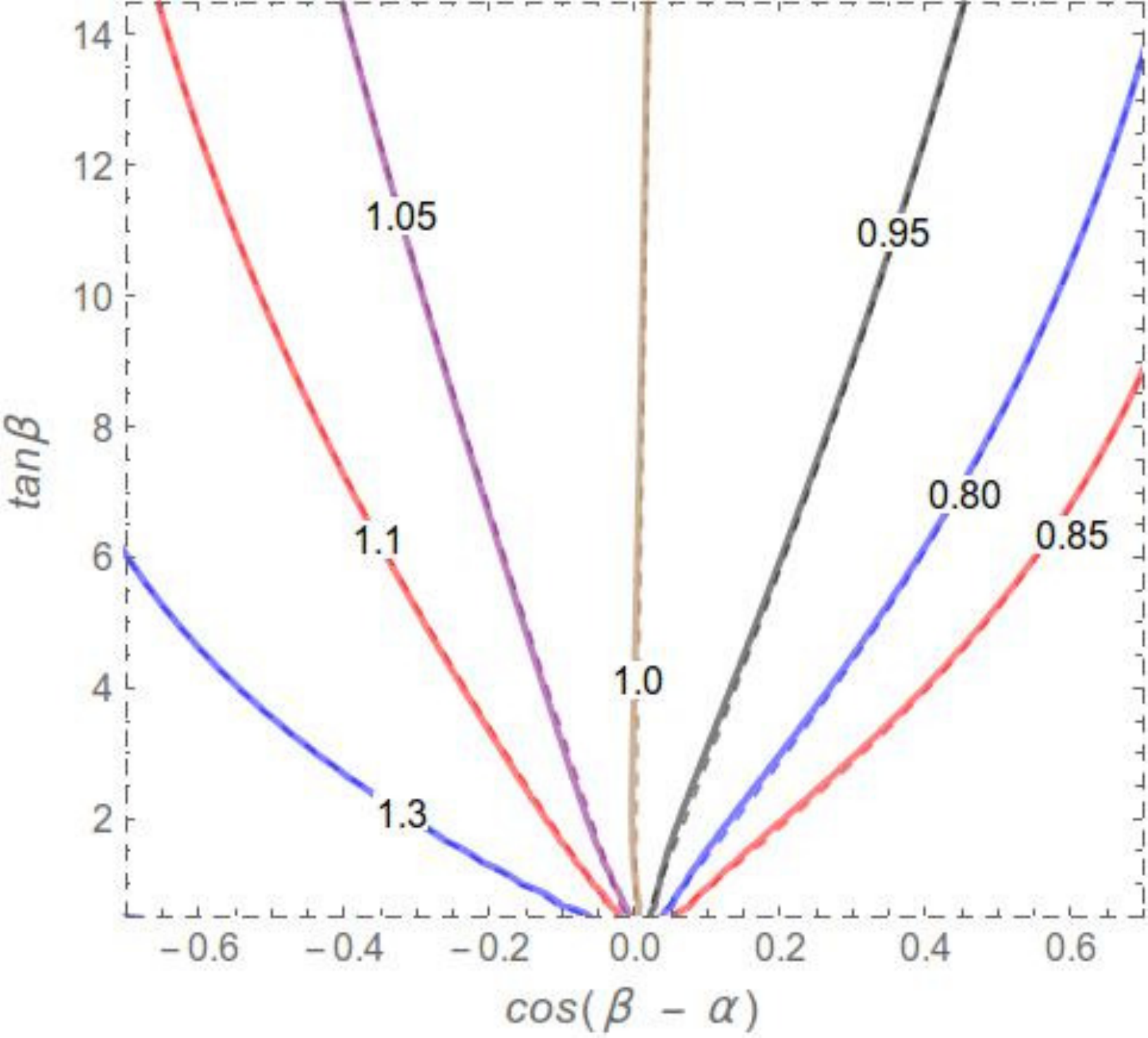}}
 \caption{(a) Dashed and solid contours represent $\sigma(gg\rightarrow h)/\sigma(gg\rightarrow h)_{SM}$ in the 2HDM at tree-level and BP1 of 2HDMEFT respectively, (b) contours of $Br(h\rightarrow VV)/Br(h\rightarrow VV)_{SM}$ in tree-level 2HDM and BP1 which almost overlap with each other.}
 \label{fig:contourbp1}
\end{center}
 \end{figure}

In order to understand the way the 6-dim operators show up in the production and decay channels separately, we show the contours of $\sigma(gg\rightarrow h)/\sigma(gg\rightarrow h)_{SM}$ and $Br(h\rightarrow VV)/Br(h\rightarrow VV)_{SM}$ for 2HDM at tree-level and for BP1 with $f = 1$~TeV in fig.~\ref{fig:contourbp1}, where $V = W,Z$.

In general, it can be seen that for positive values of the Wilson coefficients the parameter space shrinks. This happens because the maximum possible value of $\cos (\b-\a)$ is decided by the lower limit of the signal strength $\mu_{gg}^{VV}$. For positive values of the Wilson coefficients, as the coefficient of $\sin (\b-\a)$ in eqn.~(\ref{scalemult}) suggests, the effective couplings decrease. This means that in order to satisfy the lower bound on  $\mu_{gg}^{VV}$, a higher value of $\sin (\beta - \alpha)$ is required, leading to a more stringent bound on $\cos (\b-\a)$ in presence of these 6-dim operators.

To understand the shape of the allowed region on the $\cos (\b-\a)-\tan \b$ plane for $\tan \beta \gtrsim 4$, eqn.~(\ref{scalemult}) can be expanded in terms of $1/\tan \beta$, along with following simplifications which are valid up to $\mathcal{O}(1/t_{\b}^2)$: 
\bea
\label{simpl}
c_{\b} &&\rightarrow \frac{1}{t_{\b}},\hspace{7pt} s_{\b} \rightarrow 1,\hspace{7pt} c_{2\b} \rightarrow -1,\hspace{7pt} s_{2\b} \rightarrow \frac{2}{t_{\b}},\hspace{7pt}
s_{\a} \rightarrow \Big(c_{\b-\a} - \frac{s_{\b-\a}}{t_{\b}}\Big), \hspace{7pt} c_{\a} \rightarrow \Big(s_{\b-\a} +\frac{c_{\b-\a}}{t_{\b}}\Big),\nn\\
s_{2\a} &&\rightarrow 2\Big\{s_{\b-\a} c_{\b-\a} + \frac{1}{t_{\b}}(1 - 2 s_{\b-\a}^2) \Big\}, \hspace{7pt} c_{2\a} \rightarrow -1 + 2 s_{\b-\a}^2 + \frac{4 s_{\b-\a} c_{\b-\a}}{t_{\b}}.
\eea

For the values of Wilson coefficients mentioned in BP1, using relations~(\ref{simpl}) one achieves, 
\bea
\k_{hVV}^{\prime} \rightarrow  s_{\b-\a} + \frac{v^2}{f^2} \Big[ s_{\b-\a}^3 + s_{\b-\a} c_{\b-\a}^2 + \frac{1}{t_{\b}}\Big( - 2 s_{\b-\a}^2 c_{\b-\a} + \frac{1}{2} c_{\b-\a} \Big) \Big].
\eea 
For $\cos (\b-\a) \sim 0.5$, the modified coupling multiplier is always greater than $\sin (\b -\a)$. This happens due to the positive sign of the first two terms in the squared bracket. In absence of 6-dim operators, the excluded region on the positive direction of $\cos (\b-\a)$ represents the area where the value of $\mu_{gg}^{ZZ}$ is smaller than the experimentally allowed lower limit,\ie $\sim 0.76$, due to small values of $s_{\b-\a}$. For BP1, the $hVV$  coupling multiplier becomes larger than the tree-level value of $s_{\b-\a}$ in both positive and negative $c_{\b-\a}$ directions.
For instance, at $c_{\b-\a} \sim 0.5$ this leads to $\sim 15\%$  change in the decay width $\Gamma(h \rightarrow VV)$ compared to 2HDM at tree-level. Though, it can be seen from fig.~\ref{fig:contourbp1}(b), the corresponding branching ratio does not differ significantly compared to 2HDM at tree-level. This happens because the change in $\Gamma(h \rightarrow VV)$ is mostly cancelled by the same in the total decay width. However, for cross-sections no such cancellation occurs. For the case under consideration, $\sigma(ggh)/\sigma(ggh)_{SM} \sim \kappa_{htt}^2$. At $t_{\b} \sim 10$, in absence of any 6-dim terms, at $c_{\b-\a} \sim -0.37$, 
%following eqn.~(\ref{coupmulthff}), 
$\kappa_{htt} \sim 0.89$,\ie $\sigma(ggh)/\sigma(ggh)_{SM} \sim 0.79$, which leads to a decrease of $\sigma(ggh)$ by $\sim 20\%$. However, for BP1 with $f=1$~TeV, using eqn.~(\ref{hVVscalefac}) one finds that 
$\kappa_{htt}^{\prime}$($c_{\b-\a} \sim -0.47$) $\approx \kappa_{htt}(c_{\b-\a} \sim -0.37)$ and the contour of $\sigma(ggh)/\sigma(ggh)_{SM} \sim 0.8$ shifts accordingly.
As expected, the coupling multipliers in presence of higher dimensional operators do not follow the sum rules, sometimes acquiring values even greater than unity~\cite{Mrazek:2011iu}.

For BP2, using relations~(\ref{simpl}) one arrives at, 
\bea 
\k_{hVV}^{\prime} \rightarrow  s_{\b-\a} + \frac{v^2}{f^2} \Big[ -s_{\b-\a} -\frac{9}{8} c_{\b-\a} + \frac{3}{2} c_{\b-\a}^3 + \frac{1}{t_{\b}}\Big( \frac{9}{4} s_{\b-\a} - c_{\b-\a} - 3 s_{\b-\a} c_{\b-\a}^2 \Big)\Big],
\eea
which implies that in this case, for all values of $\cos (\b-\a)$, the effective coupling multiplier is smaller than the corresponding tree-level value. 
Also, the production cross-section of the SM-like Higgs decreases in this case compared to 2HDM at tree-level around $c_{\b-\a} \sim 0.45$, for which the allowed region shrinks in the direction of positive $c_{\b-\a}$. 

The non-observation of $A \rightarrow ZH (H\rightarrow b\bar{b})$ rules out  the region with low-$t_{\b}$. However, due to the change in the $Hbb$ coupling multipliers according to eqn.~(\ref{scalemult}), excluded region slightly differ in BP1 and BP2. For instance, at $c_{\b-\a} \sim -0.5$, $A \rightarrow ZH$ excludes $t_{\b} \lesssim 3.27$ for BP1 and $t_{\b} \lesssim 3.09$ for BP2. This slight change can be attributed to the modification in $Br(H \rightarrow b\bar{b})$. For instance, at $c_{\b-\a} \sim -0.5$ and $t_{\b} \sim 3$, $Br(H \rightarrow b\bar{b}) \sim 0.35$ for BP1 and $\sim 0.334$ for BP2.

\begin{figure}[h!]
 \begin{center}
 \includegraphics[width=2.8in,height=2.8in, angle=0]{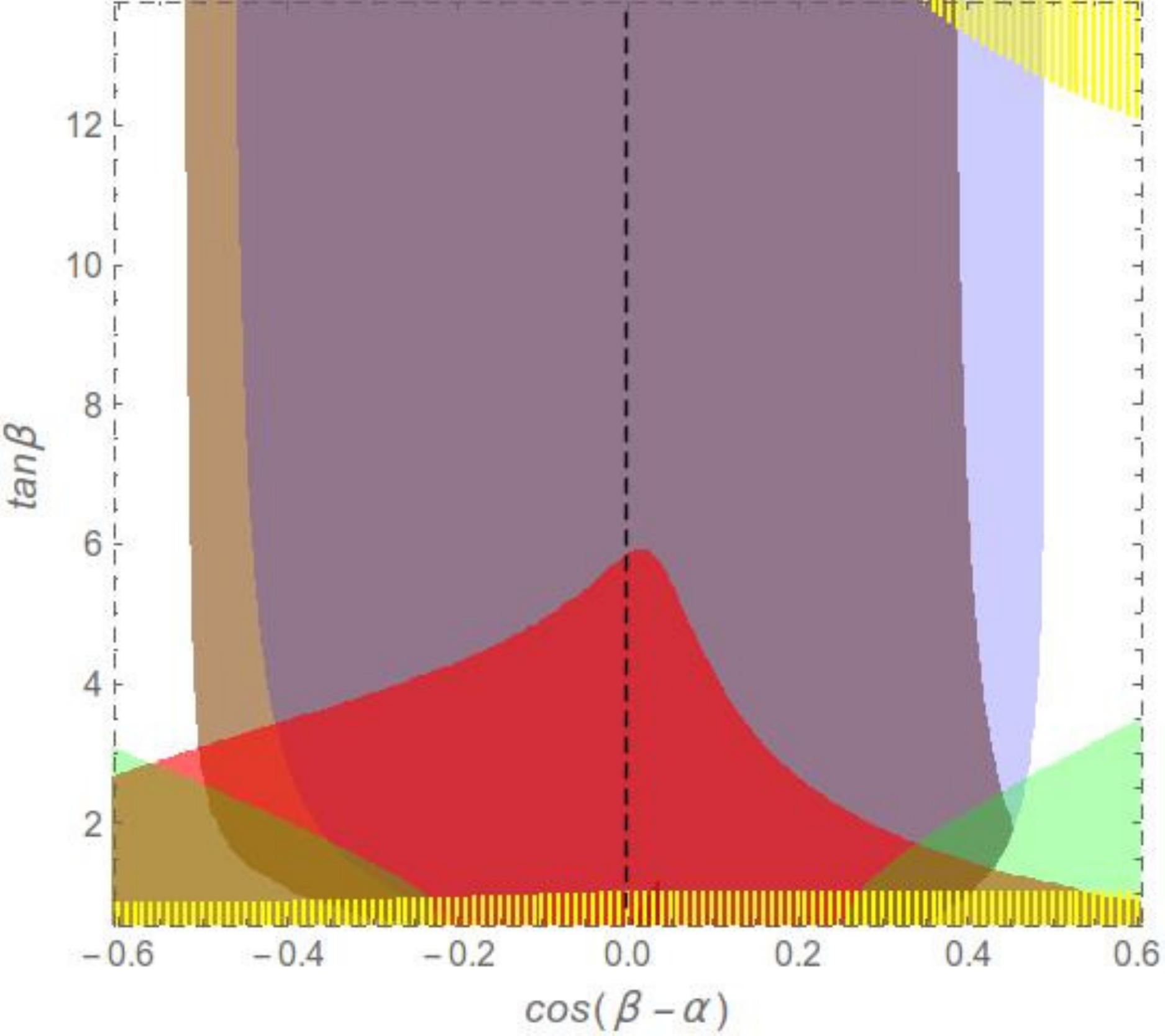}
\caption{Allowed region with a non-zero value of the Wilson coefficient of $\varphi^2 D^2 X$ type operator, $c_{\varphi W} = -0.01$, with all other Wilson coefficients being zero. Colour coding is the same as in fig. 1.}
\label{fig:2}
\end{center}
 \end{figure}

%\pagebreak

\subsubsection{${\bf \varphi^2 D^2 X}$}
These operators do not change the Higgs kinetic terms,\ie they do not rescale the couplings of the SM Higgs. But they introduce anomalous momentum structures in the $hVV$ vertices as following: 
\bea
\mathcal{L}_{hVV} &\supset & \Big(\k_{WW} W_{\m\n}^{+} W^{-\m\n} + \frac{\k_{ZZ}}{2}Z_{\m\n}Z^{\m\n} + \k_{Z\gamma} F_{\m\n}Z^{\m\n} + \frac{\k_{\gamma\gamma}}{2}F_{\m\n}F^{\m\n} \Big)\frac{h}{v}\nn\\
&&+ \Big(\k_{W\partial W} (W_{\n}^{-}D_{\m}W^{+\m\n} + h.c.) + \k_{Z\partial Z} Z_{\m} \partial_{\n} Z^{\m\n} \Big)\frac{h}{v},\nn
\eea
where the anomalous couplings can be written in terms of the Wilson coefficients of the operators in 2HDMEFT as~\cite{Karmakar:2017yek}: 
\bea
\label{phi2d2xmult}
\k_{WW} &=& -2 c_{\varphi W}, \hspace{10pt} \k_{W\partial W} = \k_{WW} - 2 c_{W}, \nn\\
\k_{ZZ} &=& \k_{W \partial W} - (2 c_{\varphi B} - 8 c_{BB})\tan^2\theta_{w}, \hspace{10pt} \k_{Z\partial Z} = \k_{W \partial W} - 2 (c_{B} + c_{\varphi B})\tan^2\theta_{w}, 
\eea
with, $c_{x} = ( -c_{\b}s_{\a} c_{x11} + s_{\b}c_{\a} c_{x22} + c_{\b-\a}c_{x12}) m_{W}^2/\Lambda^2$. $c_{xij}$, where $x = W, B, \varphi W, \varphi B, BB$, are the Wilson coefficients of the operators appearing in eqn.~(\ref{definitns}). The decay width of the SM-like Higgs into a pair of gauge bosons gets modified in the following way~\cite{Contino:2014aaa,Contino:2013kra}:  
\bea
\label{dec1}
\Gamma (h \rightarrow V^{*}V^{(*)})\bigr|_{EFT} &=& \frac{1}{\pi^2} \int_{0}^{m_h^2} \frac{dq_1^2\, \Gamma_V M_V}{(q_1^2 - M_V^2)^2 + \Gamma_V^2 M_V^2} \int_{0}^{(m_h - q_1)^2} \frac{dq_2^2 \, \Gamma_V M_V}{(q_2^2 - M_V^2)^2 + \Gamma_V^2 M_V^2}\, \Gamma (V V)\bigr|_{EFT}, \nn\\
\eea
where,
\begin{eqnarray}
\label{dec2}
\Gamma (VV)\bigr|_{EFT} &=& \Gamma (VV) \Big[1 - 2 \Big\{ \frac{a_{VV}}{2}\Big(1 - \frac{q_1^2+q_2^2}{m_{h}^2}\Big) + a_{V \partial V} \frac{q_1^2+q_2^2}{m_{h}^2}\Big\} \nn\\
&&+ a_{VV} \frac{\lambda(q_1^2,q_2^2,m_h^2)}{\lambda(q_1^2,q_2^2,m_h^2) + 12 q_1^2 q_2^2/ m_h^4}\Big(1 - \frac{q_1^2+q_2^2}{m_{h}^2}\Big)\Big],
\end{eqnarray}
with, 
\bea
\label{dec3}
a_{VV} &=& \k_{VV} \frac{m_{h}^2}{m_{V}^2}, \hspace{10pt} a_{V \partial V} = \k_{V \partial V} \frac{m_{h}^2}{2 m_{V}^2}, \nn\\
\Gamma (VV) &=& \sin^2 (\b - \a)\frac{\delta_V G_{F} m_h^3}{16 \sqrt{2} \pi} \sqrt{\lambda(q_1^2,q_2^2,m_h^2)}\Big(\lambda(q_1^2,q_2^2,m_h^2) + \frac{12 q_1^2 q_2^2}{m_h^4}\Big),
\eea
with $\delta_V = 2,1$ for $V = W,Z$ respectively, and $\lambda(x,y,z) = (1 - x/z - y/z)^2 - 4xy/z^2$. 

This kind of operators contributes to the precision observables and various cross sections related to the SM-like Higgs, which in turn put bounds on the coefficients of these operators. Operators $O_{Wij}, O_{Bij}, O_{\varphi Bij}$ are constrained at $\sim \mathcal{O}(10^{-3})$. However, the Wilson coefficients of $O_{\varphi W ij}$ are constrained only at $\sim \mathcal{O}(10^{-2})$ in the negative direction~\cite{Karmakar:2017yek,Pomarol:2013zra,Falkowski:2013dza}. This fact is also reflected when we see the impact of such operators on the alignment limit. All the operators of this kind, except $O_{\varphi W ij}$, are constrained enough to inflict any changes on the $\cos (\beta - \alpha) -\tan \beta$ plane compared to tree-level 2HDM. We show the effect of these operators along with the value of the combination of Wilson coefficients $c_{\varphi W} = -0.01$ in fig.~\ref{fig:2}. In this case, the minimum allowed value of $\cos (\beta - \alpha)$  in the negative direction changes from $-0.39$ to $-0.49$ at $\tan \beta \sim 3.7$,\ie increases by $+25\%$.

\noindent  \subsection{{\bf Type-II 2HDM}}

\begin{figure}[h!]
 \begin{center}
\subfigure[]{
 \includegraphics[width=2.8in,height=2.8in, angle=0]{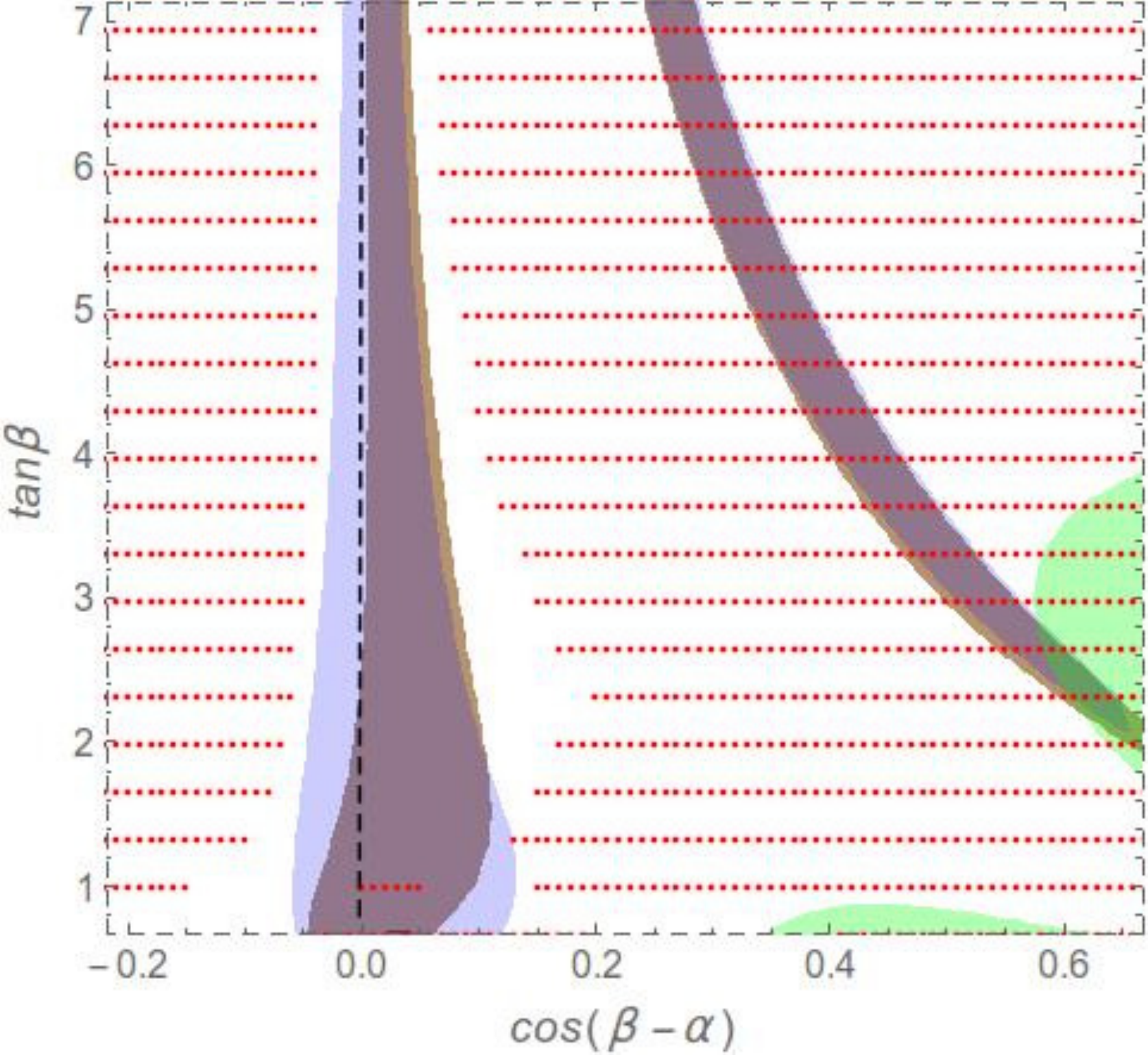}}
 \hskip 15pt
 \subfigure[]{
 \includegraphics[width=2.8in,height=2.8in, angle=0]{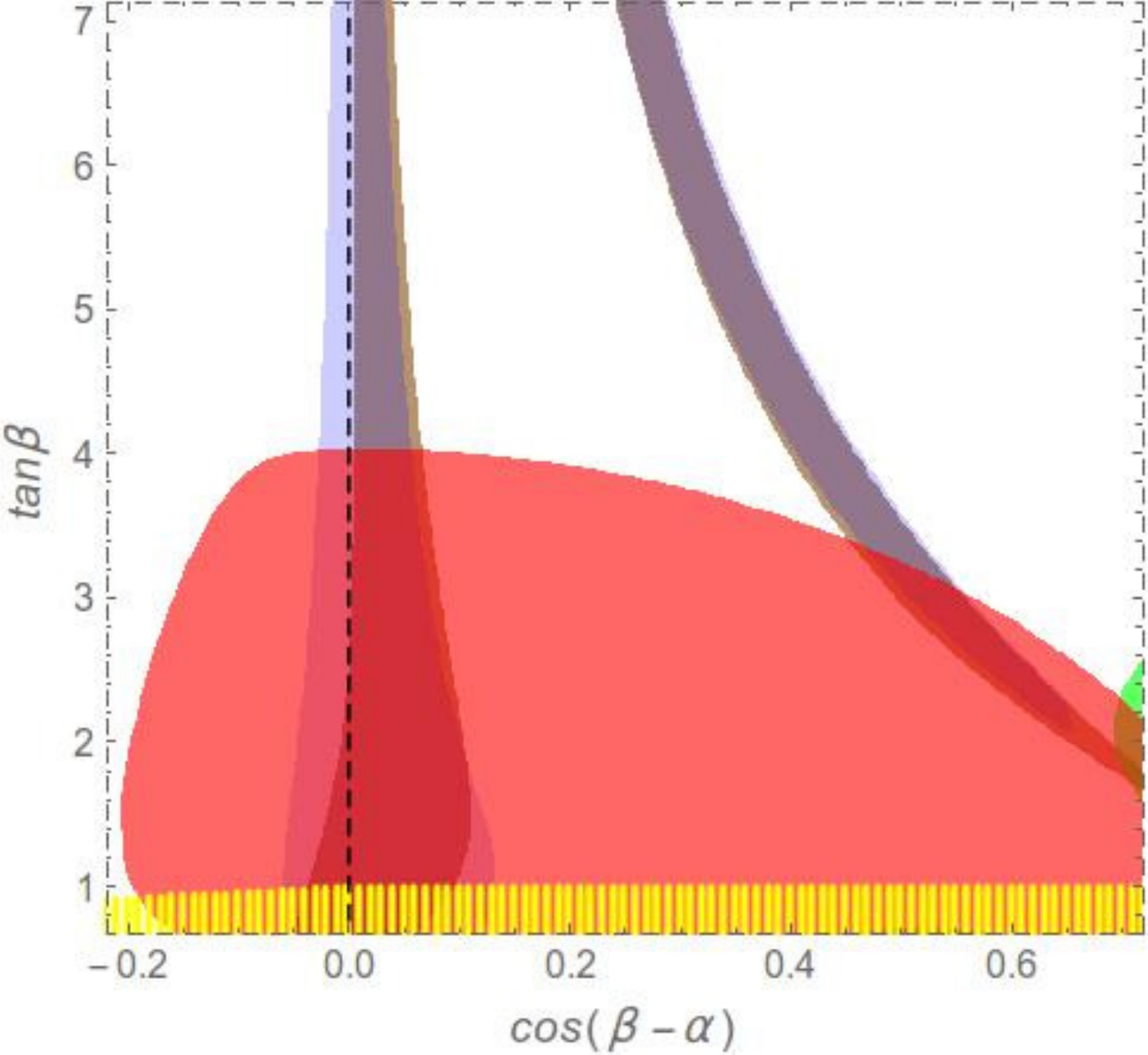}}
 \caption{Modification of alignment limit in BP3 for Type-II 2HDM. (a) $m_H \sim 415$ GeV, $m_A \sim 480$ GeV, $m_{H^{\pm}} \sim 485$ GeV. Red dotted area violate theoretical constraints, rest of the colour coding is the same as in fig. 1. (b) $m_H \sim 150$ GeV, and $m_A \sim m_{H^{\pm}} \sim 485$ GeV. Solid red region is ruled out from $A \rightarrow ZH$. Rest of the colour coding is the same as in fig. 1.}
 \label{fig:5}
\end{center}
\end{figure}

As mentioned earlier, the allowed region on the $\cos (\b-\a)-\tan\b$ plane for Type-II 2HDM is much smaller than the same for Type-I 2HDM. For Type-II 2HDM, we have checked that the operators of type $\varphi^2 D^2 X$ and $\varphi^2 X^2$ do not change the shape of the allowed region on the $\cos (\b-\a)-\tan \b$ plane significantly, because the Wilson coefficients are too small in size, after considering bounds from precision measurements. However, the $\varphi^4 D^2$ type of operators can lead to considerable changes in the allowed range of $\cos (\b -\a)$ for $\tan \beta \sim 1 - 10$. We illustrate one such case in fig.~\ref{fig:5}(a) for which the values of relevant parameters are given by: 

\begin{itemize}
\item {\bf BP3} \hspace{5pt} $c_{H1} = -1$, $c_{H2} =  \frac{3}{2}$, $c_{H12} = c_{H1H2} =  c_{H1H12} = c_{H2H12} = 0$, $f = 1$ TeV,
\end{itemize} 
with the following values of the 2HDM parameters:

$m_{h} = 125.7$ GeV, $m_H = 415$ GeV, $m_A = 480$ GeV, $m_{H^{\pm}} = 485$ GeV, $\mu^2 = \mu^2_{*}$. 

In  fig.~\ref{fig:5}(a), at $t_\b \sim 4$, there is $+21\%$ change in positive $c_{\b-\a}$ direction as the maximum allowed value of $c_{\b-\a}$ goes to $0.066$ from $0.05$. In the  negative $c_{\b-\a}$ direction, the minimum value changes from $-0.0267$ to $0.0055$,\ie increases by $120\%$. For higher $t_{\b}$ this deviation can become larger. An interesting effect for this benchmark scenario is, for most of the values of $t_{\b}$, $c_{\b-\a} = 0$ remains excluded from the measurement of Higgs signal strengths at $95\%$ CL. In this case, the bounds from the non-observation of $H$ through various decay channels, not even $H \rightarrow hh$, affect the relevant parameter space on the $\cos (\b -\a) - \tan \beta$ plane. The constraint from $b \rightarrow s \gamma$ is taken care of by considering $m_{H^{\pm}} > 480$ GeV. Though the constraints from $A \rightarrow Z h$ are non-zero, those do not significantly overlap with the region allowed from $h \rightarrow VV^{*}$. 

In case of relatively lower masses of the heavier $CP$-even neutral scalar $H$, regions up to $\tan \b \sim 5-6$ can be ruled out from the non-observation of $A \rightarrow ZH$~\cite{Dorsch:2016tab}. In such cases, the theoretical bounds get comparatively relaxed, but the effects of 6-dim operators on the $\cos (\b-\a)-\tan \b$ plane still remain significant. We present one such scenario with the exotic scalar masses being, $m_H \sim 150$ GeV, and $m_A \sim m_{H^{\pm}} \sim 485$ GeV, and all the other parameters having values the same as in case of BP3. This particular case is depicted in fig.~\ref{fig:5}(b).

\begin{figure}[h!]
 \begin{center}
\subfigure[]{
 \includegraphics[width=2.8in,height=2.8in, angle=0]{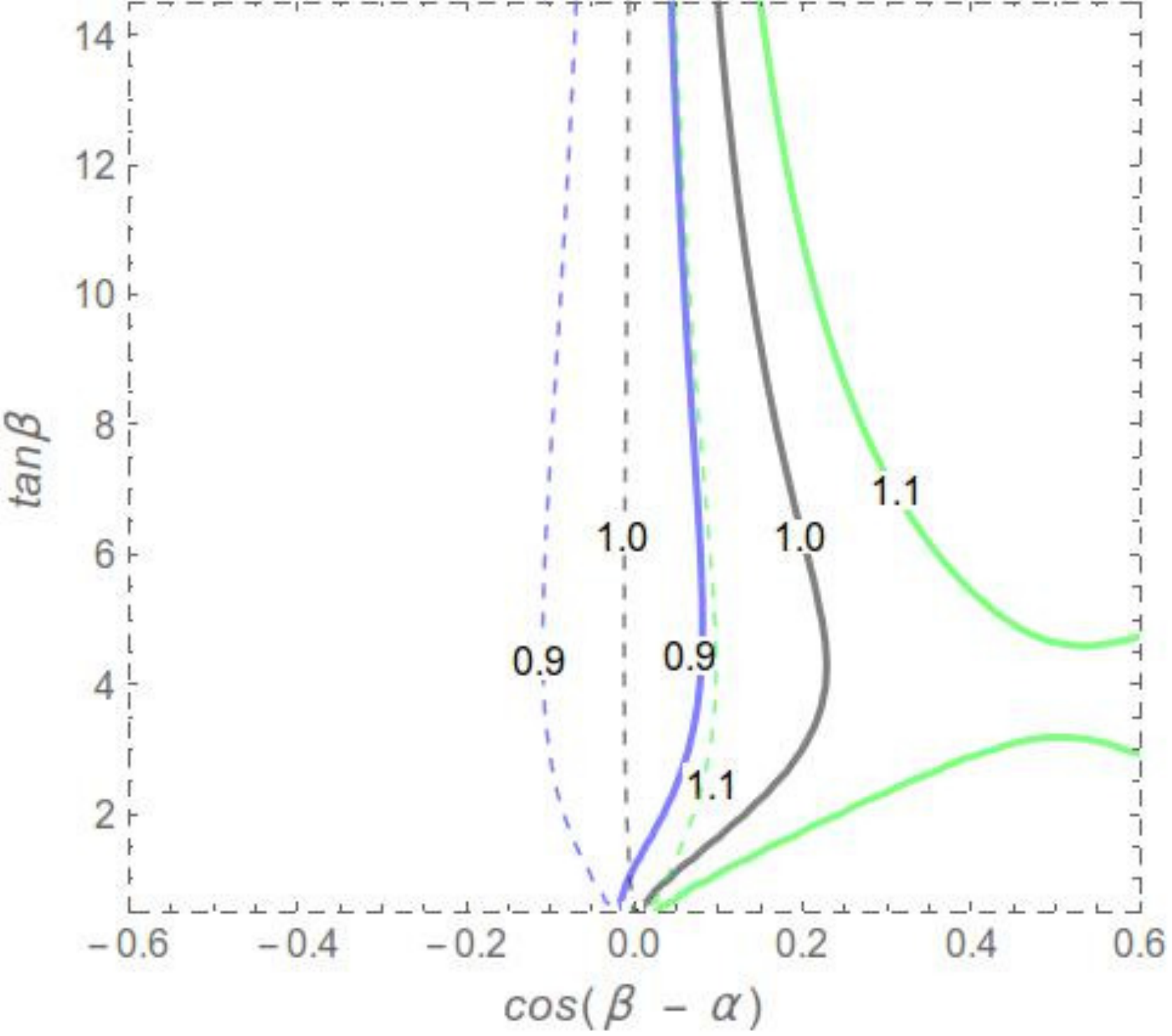}}
 \hskip 15pt
 \subfigure[]{
 \includegraphics[width=2.8in,height=2.8in, angle=0]{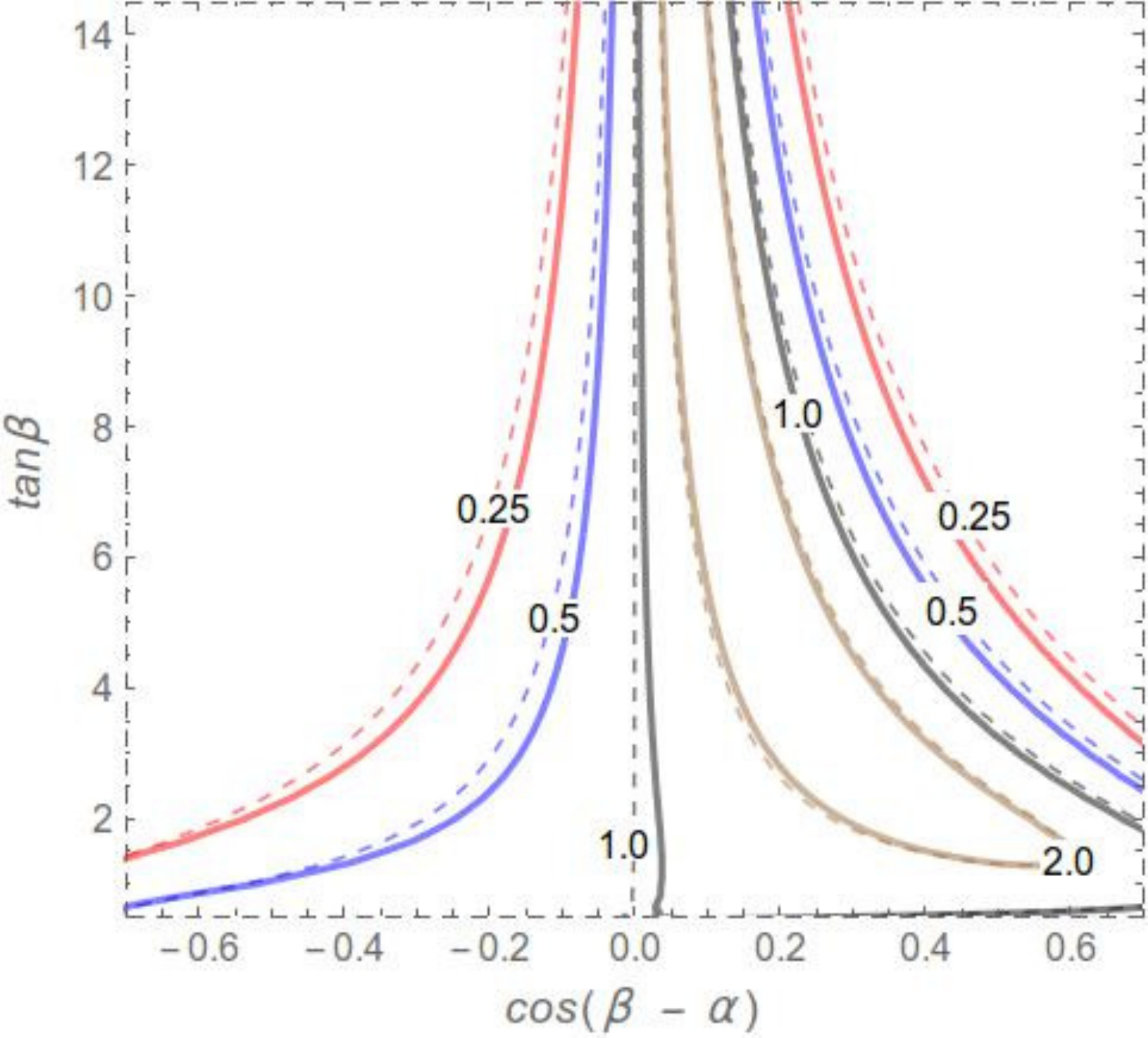}}
 \caption{
Dashed and solid lines represent respectively for 2HDM at tree-level and for BP3 of 2HDMEFT, contours of
 %For BP3, (a) contours of 
 (a) $\sigma(gg\rightarrow h)/\sigma(gg\rightarrow h)_{SM}$, (b) $Br(h\rightarrow VV)/Br(h\rightarrow VV)_{SM}$.}
 \label{fig:contourbp3}
\end{center}
\end{figure}

For Type-II 2HDM, in the positive $\cos (\b-\a)$ direction the allowed region around $\cos (\b-\a) \sim 0$ represents the area where, $\mu_{gg}^{VV} \lesssim 1.38$, whereas on the negative direction it represents $\mu_{gg}^{VV} \gtrsim 0.76$. For BP3, the coupling multiplier approximately assumes the form, $\k_{hVV}^{\prime} \sim (s_{\b-\a} - 1.5 v^2 s_{\b -\a}^3/f^2)$ for vanishing $\cos (\b-\a)$. This implies that for both positive and negative small values of $\cos (\b-\a)$, the coupling multiplier in presence of 6-dim operators is less than that of the tree-level counterpart. 
We have shown the contours of $\sigma(gg\rightarrow h)/\sigma(gg\rightarrow h)_{SM}$ and $Br(h\rightarrow VV)/Br(h\rightarrow VV)_{SM}$ for type-II 2HDM at tree-level and for BP3 of 2HDMEFT in fig.~\ref{fig:contourbp3}. Both the ratios for BP3 shown in fig.~\ref{fig:contourbp3}~(a) and (b) take unit values at some positive value of $\cos (\b-\a)$, leading to an overall positive shift in the allowed parameter space compared to tree-level 2HDM.

At this point it is important to mention that for both Type-I and II 2HDM, though the allowed range of $\cos (\b-\a)$ changes significantly, the coupling multipliers of the SM-like Higgs boson at the tree-level still has to be quite close to unity.  The reason for this is, as eqns.~(\ref{scalemult}) and (\ref{hVVscalefac}) together suggest, the deviation in the coupling multipliers are all functions of $\cos (\b-\a)$, $\sin \b$ and $\cos \b$ when $\varphi^4 D^2$ type of operators are added in the Lagrangian. Eqn.~(\ref{phi2d2xmult}) suggests that the situation is similar when $\varphi^2 D^2 X$ type of operators are added. Due to such dependence on $\b$ and $\a$, the deviations in coupling multipliers are never drastic, $ |\delta \kappa | \lesssim 0.5$. 

\noindent  \subsection{{\bf A concrete example: UV-friendly Little Higgs model}}
We have seen earlier in this section that new physics scenarios beyond 2HDM can change the allowed region of 2HDM parameter space in terms of $\cos (\beta - \alpha)$. In order to show the effect of these operators for a concrete model, we consider the so-called `UV-friendly' Little Higgs model without $T$-parity which is based on the coset $SU(6)/Sp(6)$~\cite{Brown:2010ke}, along with Type-I Yukawa couplings. This model, at low energies, can be parametrised by the following values of the Wilson coefficients of the $\varphi^4 D^2$ operators~\cite{Kikuta:2012tf}: 
\bea
\label{littleh}
c_{H1} = c_{H2} = 2, \hspace{10pt} c_{H1H2} = \frac{1}{2}, \hspace{10pt}
c_{H12} = -\frac{3}{2}, \hspace{10pt} c_{H1H12} = c_{H2H12} = 0. 
\eea 
The allowed parameter space of this model in the $\cos (\beta - \alpha) -\tan \beta$ plane with $f = 1.4$ TeV is shown in fig.~6. Such values of the NP scale $f$ are still allowed from 13~TeV LHC data for this Little Higgs model~\cite{Dercks:2018hgz}. In this case the modified $hVV$ coupling multiplier assumes the form: 
\bea
\k_{hVV}^{\prime} \rightarrow s_{\b-\a} +\frac{v^2}{f^2}\Big(- 2 s_{\b-\a}^3 - 2 s_{\b-\a}^2 c_{\b -\a} +  \frac{1}{t_{\b}} (0.625 c_{\b -\a} - 6 s_{\b-\a}^2 c_{\b -\a} - 3.5  c_{\b -\a}^3 ) \Big).\nn
\eea
This implies that for positive values of $c_{\b -\a}$, the modified coupling multiplier decreases compared to the same in 2HDM. Along with that, the production cross-section also decreases in both negative and positive directions, thus shrinking the allowed region on the $c_{\b-\a} - t_{\b}$ plane from both directions.
For a composite 2HDM based on coset $SO(6)/SO(4)\times SO(2)$~\cite{DeCurtis:2016tsm,DeCurtis:2017gzi,DeCurtis:2016scv} the allowed range of $\cos (\b-\a)$ shrinks as well.

\begin{figure}[h!]
\begin{center}
 \includegraphics[width=2.8in,height=2.8in, angle=0]{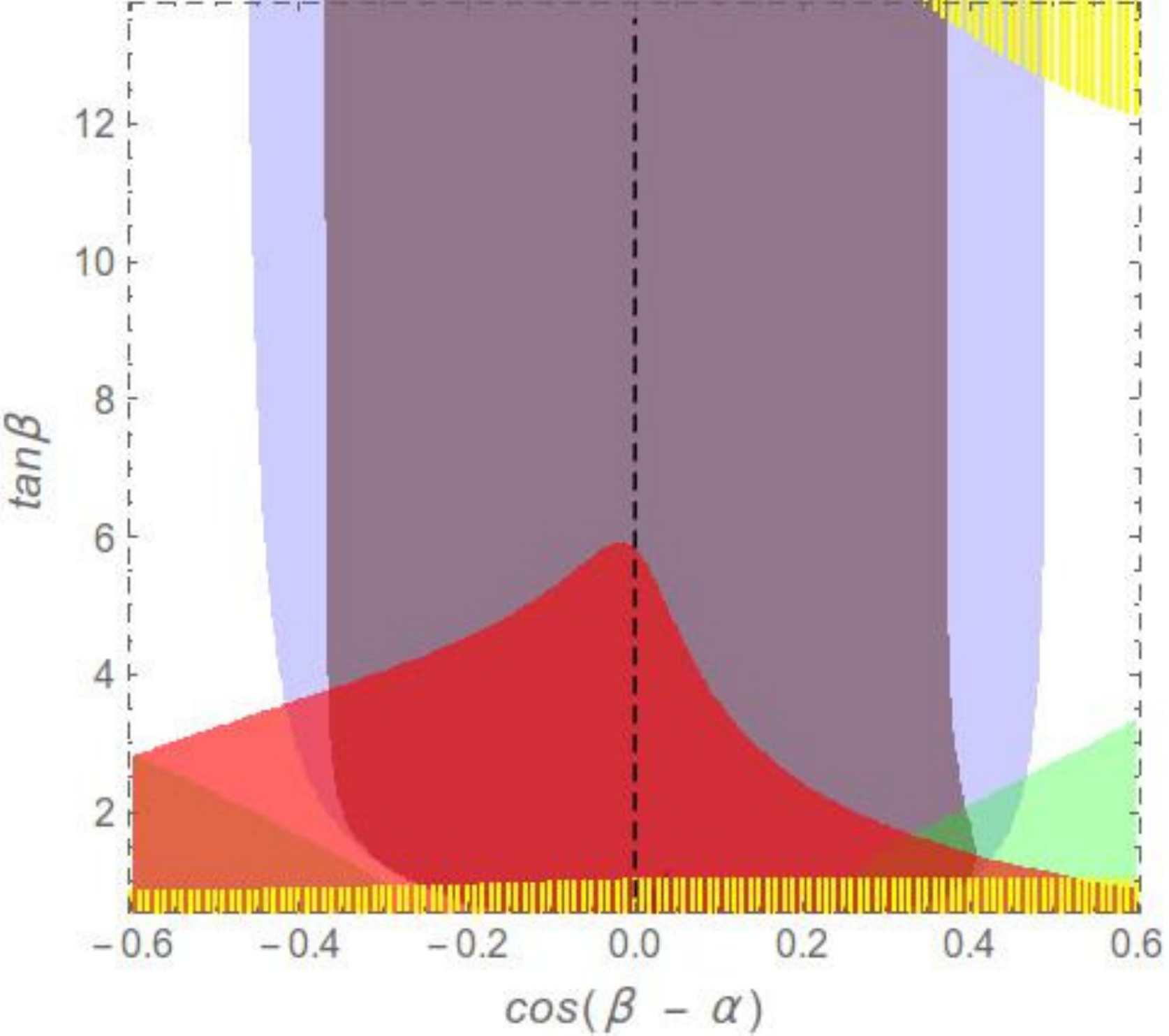}
\label{fig:uvcomp}
\caption{ Little Higgs model based on $SU(6)/Sp(6)$. Colour coding is the same as 
in fig. 1. }
\end{center}
\end{figure}

\section{Summary and Discussion}
\label{summary}
The discovery of the SM-like Higgs boson and the compatibility of the measured signal strengths with the SM-predicted values put stringent restrictions on any model with an extended scalar sector. A 2HDM is then realised close to the `alignment limit'.  The degree of alignment in a 2HDM  depends on the type of Yukawa couplings. For example, the Type-II 2HDM is restricted to be more aligned compared to Type-I 2HDM. An exact alignment in 2HDM can point to the existence of an underlying symmetry. Hence, it is important to scrutinise the robustness of the method of extraction of the degree of alignment in 2HDM. 

In this paper, we argue that as the experiments only constrain the effective coupling of the SM-like Higgs boson with other SM particles, we are restricting only the `effective alignment' rather than the `true alignment' in a 2HDM.
These two alignments differ in presence of physics beyond a 2HDM,  which can be encoded in the language of 2HDMEFT. SMEFT is not adequate for such a study as the exotic scalar particles can be light enough so that they are not decoupled from the mass spectrum of the low-energy theory.

The theoretical prediction of the signal strengths of the SM-like Higgs boson can be sensitive to the 6-dim operators in 2HDMEFT as they can alter the production cross-section and the decay width of this Higgs. When confronted with the measurements from ATLAS and CMS collaborations, the presence of such operators do result in changing the allowed parameter space. In this article, we consider only the bosonic operators and choose the $\cos (\beta - \alpha) - \tan \beta$ plane of the 2HDM parameter space to display the impact of such operators on `true' 2HDM alignment.

Fortunately, most of the bosonic operators are not significant as far as alignment limit is concerned. For example, the $\varphi^6$ operators are not important because their effects are negligible in the decay of SM-like Higgs boson. Some of the operators of the $\varphi^4 D^2$ category contribute to the $T$-parameter at the tree-level. So the corresponding Wilson coefficients are constrained at the per-mille level and their contributions to Higgs phenomenology are rather small. Most of the operators of type $\varphi^2 D^2 X$ and $\varphi^2 X^2$ are also not interesting due to the same reason, as they are constrained at $\sim \mathcal{O}(10^{-3})$ from the electroweak precision tests. However, a few operators of type $\varphi^2 D^2 X$ are constrained only at $\sim \mathcal{O}(10^{-2})$ and can be significant due to their specific Lorentz structures. Also, some of the operators of type $\varphi^4 D^2$, those which do not contribute to the $T$-parameter, can lead to substantial deviation in the production and decay width of the SM-like Higgs boson. In passing, we reiterate that the sum rules involving the couplings of the CP-even neutral scalars to the $W$ and $Z$ bosons do not hold if the tree-level 2HDM is extended with the 6-dim operators of type $\varphi^4 D^2$.

The difference between the `effective' and `true' alignments is sensitive to the Wilson coefficients of these 6-dim operators, the 2HDM parameters and the type of 2HDM considered. Depending on these, the allowed parameter space on the  $\cos (\beta - \alpha) - \tan \beta$ plane can shift, shrink and what is even more interesting, the exact alignment limit can be excluded. We have demonstrated this by the choice of suitable benchmark points~\cite{Banerjee:2017wmg}. It is noticed that for Type-I 2HDM, the region allowed from measurements of the signal strengths of the SM-like Higgs boson being quite larger compared to other variants of 2HDM, the 6-dim operators are able to inflict substantial changes, often increasing or decreasing the allowed value of $\cos(\beta -\alpha)$ by $\sim \mathcal{O}(0.1)$. In case of Type-II 2HDM, the percentage change in the allowed value of $\cos(\beta -\alpha)$ can be larger than  Type-I 2HDM, achieving values $\sim 100\%$ or higher, whereas such changes for Type-I 2DHM reaches values up to $\sim 25\%$. We have seen that, negative values of the Wilson coefficients lead to a larger allowed range of $\cos (\beta - \alpha)$ in both positive and negative directions in most of the cases. We also studied a particular Little Higgs model with Type-I Yukawa coupling as a UV-complete example of 2HDMEFT. Generally the impact of $\varphi^2 D^2 X$ type of operators on 2HDM alignment is smaller compared to the $\varphi^4 D^2$ operators. In Type-II 2HDM, it is also noticed that in the presence of the 6-dim operators, the exact alignment limit, $\cos (\b-\a) = 0$, can be ruled out for a wide range of values of $\tan \b$ at $95\%$ CL. All these demonstrate that the 6-dim operators in 2HDMEFT are capable of masking the true alignment.

\section{Acknowledgements}
S.K. thanks A. Falkowski and J. M. No for email communications. S.K. acknowledges T. Stefaniak for help regarding {\tt 2HDMC}. Authors would also like to thank P. S. Bhupal Dev for discussions. This work is supported by the Department of Science and Technology via Grant Nos. EMR/2014/001177 and INT/FRG/DAAD/P-22/2018. Authors also thank the referee for valuable comments.

\setcounter{footnote}{0}
\renewcommand*{\thefootnote}{\arabic{footnote}}

%==============================================================================
%==============================================================================

\appendix
\section{Scalar field redefinition and couplings}
\label{appendix1}
The redefinition of Higgs fields is given by eqn.~(\ref{hfieldred}), 
where,
\bea
\label{xexp}
x_1 &=& \frac{s_{\a}^2 \Delta_{11\rho}}{4f^2} + \frac{c_{\a}^2 \Delta_{22\rho}}{4f^2} - \frac{s_{\a} c_{\a} \Delta_{12\rho}}{4f^2},\nn\\
x_2 &=& \frac{c_{\a}^2 \Delta_{11\rho}}{4f^2} + \frac{s_{\a}^2 \Delta_{22\rho}}{4f^2} + \frac{s_{\a} c_{\a} \Delta_{12\rho}}{4f^2},\nn\\
y &=& \frac{c_{\a} (\Delta_{11\rho} - \Delta_{22\rho})}{4f^2} - \frac{c_{2\a}  \Delta_{12\rho}}{4f^2}.
\eea
Along with eqns.~(\ref{xexp}) and the following expressions,  
\bea
\label{deltas} 
\Delta_{11\rho} &=& 4 c_{H1} v_1^2 + 4 c_{H12} v_2^2 + 4c_{H1H12} v_1 v_2, \nn\\
\Delta_{22\rho} &=& 4 c_{H12} v_1^2 + 4 c_{H2} v_2^2 + 4c_{H2H12} v_1 v_2, \nn\\
\Delta_{12\rho} &=& 2 c_{H1H12} v_1^2 + 2 c_{H2H12} v_2^2 + \Big(4 c_{H12} + 2 c_{H1H2}\Big) v_1 v_2,
\eea
one gets,
\bea
\label{hVVscalefac}
x_1 &=&\frac{v^2}{f^2} \Big( c_{H1} c_{\b}^2  s_{\a}^2 + c_{H2} c_{\a}^2 s_{\b}^2 + \frac{1}{8} c_{H1H2} s_{2\a} s_{2\b} + c_{H12} (c_{\a}^2 c_{\b}^2 + s_{\a}^2 s_{\b}^2 - \frac{1}{4} s_{2\a} s_{2\b})\nn\\
&&\hspace{50pt}+ c_{H1H12} c_{\b} s_{\a} (s_{\a} s_{\b} - \frac{1}{2} c_{\a} c_{\b}) + c_{H2H12} c_{\a} s_{\b} (c_{\a} c_{\b} - \frac{1}{2} s_{\a} s_{\b}) \Big) ,\nn\\
x_2 &=&\frac{v^2}{f^2} \Big( c_{H1} c_{\b}^2 c_{\a}^2 + c_{H2} s_{\a}^2 s_{\b}^2 + \frac{1}{8} c_{H1H2} s_{2\a} s_{2\b} + c_{H12} (s_{\a}^2 c_{\b}^2 + c_{\a}^2 s_{\b}^2 - \frac{1}{4} s_{2\a} s_{2\b})\nn\\
&&\hspace{50pt}+ c_{H1H12} c_{\b} c_{\a} (c_{\a} s_{\b} - \frac{1}{2} s_{\a} c_{\b}) + c_{H2H12} s_{\a} s_{\b} (s_{\a} c_{\b} - \frac{1}{2} c_{\a} s_{\b}) \Big),\nn\\
y &=& \frac{v^2}{f^2} \Big( \frac{1}{2} c_{H1} s_{2\a} c_{\b}^2 -\frac{1}{2} c_{H2}  s_{2\a} s_{\b}^2 - \frac{1}{8} c_{H1H2} c_{2\a} s_{2\b} - \frac{1}{2} c_{H12} (  c_{2\b}  s_{2\a} + \frac{1}{2} c_{2\a}  s_{2\b}) \nn\\
&&\hspace{50pt}+ \frac{1}{4} c_{H1H12} ( s_{2\a} s_{2\b} - c_{2\a} c_{\b}^2 ) - \frac{1}{4} c_{H2H12} ( s_{2\a} s_{2\b} + c_{2\a} s_{\b}^2 )\Big).
\eea

Eqns.~(\ref{hVVscalefac}) along with eqns.~(\ref{scalemult}) give us the modified coupling multipliers for both the CP-even neutral Higgs bosons in case of 2HDMEFT.

\end{document}